\documentclass[12pt,letterpaper,english,aps,superscriptaddress,groupedaddress,showpacs,showkeys]{revtex4}
\usepackage[T1]{fontenc}
\usepackage[latin9]{inputenc}
\usepackage{babel}
\usepackage{rotating}
\usepackage{calc}
\usepackage{amsmath}
\usepackage{amssymb}
\usepackage[unicode=true]
 {hyperref}
\usepackage{breakurl}

\makeatletter

\providecommand{\tabularnewline}{\\}

\@ifundefined{textcolor}{}
{%
 \definecolor{BLACK}{gray}{0}
 \definecolor{WHITE}{gray}{1}
 \definecolor{RED}{rgb}{1,0,0}
 \definecolor{GREEN}{rgb}{0,1,0}
 \definecolor{BLUE}{rgb}{0,0,1}
 \definecolor{CYAN}{cmyk}{1,0,0,0}
 \definecolor{MAGENTA}{cmyk}{0,1,0,0}
 \definecolor{YELLOW}{cmyk}{0,0,1,0}
}

\usepackage{dcolumn}
\usepackage{bm}
\usepackage{amsfonts}

\makeatother

\begin{document}

\title{On wave equations for the Majorana particle in (3+1) and (1+1) dimensions}

\author{Salvatore De Vincenzo}

\homepage{https://orcid.org/0000-0002-5009-053X}

\email{[salvatore.devincenzo@ucv.ve]}

\selectlanguage{english}%

\affiliation{Escuela de F\'{\i}sica, Facultad de Ciencias, Universidad Central de Venezuela,
A.P. 47145, Caracas 1041-A, Venezuela.}

\thanks{I would like to dedicate this paper to the memory of my beloved father
Carmine De Vincenzo Di Fresca, who passed away unexpectedly on March
16, 2018. That day something inside of me also died.}

\date{January 17, 2021}

\begin{abstract}
\noindent \textbf{Abstract} In general, the relativistic wave equation
considered to mathematically describe the so-called Majorana particle
is the Dirac equation with a real Lorentz scalar potential plus the
so-called Majorana condition. Certainly, depending on the representation
that one uses, the resulting differential equation changes. It could
be a real or a complex system of coupled equations, or it could even
be a single complex equation for a single component of the entire
wave function. Any of these equations or systems of equations could
be referred to as a Majorana equation or Majorana system of equations
because it can be used to describe the Majorana particle. For example,
in the Weyl representation, in (3+1) dimensions, we can have two non-equivalent
explicitly covariant complex first-order equations; in contrast, in
(1+1) dimensions, we have a complex system of coupled equations. In
any case, whichever equation or system of equations is used, the wave
function that describes the Majorana particle in (3+1) or (1+1) dimensions
is determined by four or two real quantities. The aim of this paper
is to study and discuss all these issues from an algebraic point of
view, highlighting the similarities and differences that arise between
these equations in the cases of (3+1) and (1+1) dimensions in the
Dirac, Weyl, and Majorana representations. Additionally, to reinforce
this task, we rederive and use results that come from a procedure
already introduced by Case to obtain a two-component Majorana equation
in (3+1) dimensions. Likewise, we introduce for the first time a somewhat
analogous procedure in (1+1) dimensions and then use the results we
obtain.
\end{abstract}

\pacs{03.65.-w, 03.65.Ca, 03.65.Pm}

\keywords{relativistic quantum mechanics of a single-particle; the Dirac equation
and the equations for the Majorana particle; the Dirac, Weyl, and
Majorana representations.}

\maketitle

\section{Introduction}

\noindent In general, the relativistic wave equation considered to
mathematically describe a first quantized Majorana particle (an electrically
neutral fermion in (3+1) dimensions that is its own antiparticle)
is the Dirac equation with a real Lorentz scalar potential together
with the so-called Majorana condition \cite{RefA,RefB}. The latter
condition assumes that the Dirac wave function is equal to its respective
charge-conjugate wave function, i.e., $\Psi=\Psi_{C}$; this is regardless
of the representation of the gamma matrices that one chooses when
writing the Dirac equation. This way of characterizing the Majorana
particle can be implemented in (3+1) dimensions and also in (1+1)
dimensions, although in the latter case one would be describing  the
one-dimensional Majorana particle. 

As might be expected, depending on the representation that one uses
when writing the Dirac equation and the Majorana condition, and without
distinguishing between (3+1) and (1+1) dimensions, the differential
equation that one obtains changes. It could be a real or a complex
system of coupled equations or even a single complex equation for
a single component of the entire wave function and whose solution,
together with the relation that emerges from the Majorana condition,
would allow one to build the entire wave function \cite{RefC,RefD,RefE,RefF}.
Certainly, any of these equations or systems of equations could be
referred to as a Majorana equation or Majorana system of equations
because any of them could be used to describe the Majorana particle. 

Unexpectedly, the equation generally known in the literature as the
Majorana equation is a relativistic wave equation similar to the free
Dirac equation, $\mathrm{i}\hat{\gamma}^{\mu}\partial_{\mu}\Psi-\frac{\mathrm{m}c}{\hbar}\hat{1}\Psi=0$
($\hat{1}$ is the identity matrix, which is a $4\times4$ matrix
in (3+1) dimensions but a $2\times2$ matrix in (1+1) dimensions),
but in addition to the Dirac wave function $\Psi$, the Majorana equation
also includes the respective charge-conjugate wave function $\Psi_{C}$.
The equation in question is usually written as $\mathrm{i}\hat{\gamma}^{\mu}\partial_{\mu}\Psi-\frac{\mathrm{m}c}{\hbar}\hat{1}\Psi_{C}=0$
\cite{RefG}, and by using typical properties associated with the
charge conjugation operation, one obtains $\mathrm{i}\hat{\gamma}^{\mu}\partial_{\mu}\Psi_{C}-\frac{\mathrm{m}c}{\hbar}\hat{1}\Psi=0$;
both of these equations imply that $\Psi$ and $\Psi_{C}$ satisfy
the well-known Klein-Fock-Gordon equation. In writing the Majorana
equation, it is important to remember that $\Psi_{C}$ has the same
transformation properties as $\Psi$ under proper Lorentz transformations;
hence, this equation is Lorentz covariant. Likewise, the Majorana
condition is Lorentz covariant \cite{RefC}. The Majorana equation
could describe hypothetical particles that have been called Majoranons
\cite{RefG}. Clearly, the Majorana equation together with the Majorana
condition can also lead to equations for the Majorana particle \cite{RefA}.
It can be stated that the uncharged Majorana particle (for which $\Psi=\Psi_{C}$
is satisfied) would be the physical solution, and the charged Majoranon
(for which the Majorana condition is not imposed) would be the unphysical
solution of the Majorana equation \cite{RefG}. We wish to point out
in passing that the Majorana equation can also admit a Lorentz scalar
potential. 

In general, when characterizing a Majorana particle with the help
of complex four-component wave functions (in (3+1) dimensions) or
two-component wave functions (in (1+1) dimensions) these components
are not all independent because the Majorana condition must be satisfied.
Apropos of this, in the Majorana representation, the Majorana condition
becomes the reality condition of the wave function, i.e., $\Psi=\Psi^{*}$;
therefore, we can conclude that in (3+1) or (1+1) dimensions, the
wave function that describes the Majorana particle has four or two
independent real components, and these real components can be accommodated
just in two or one independent complex components or component \cite{RefC}.
Then, to describe the Majorana particle in (3+1) or (1+1) dimensions,
a four-component or two-component wave function is not absolutely
necessary, i.e., a four-component or two-component scheme or formalism
is not absolutely necessary; it can also be done with two-component
or one-component wave functions, i.e., a two-component or one-component
scheme or formalism in (3+1) or (1+1) dimensions is sufficient. 

Returning to the issue of the equations for the Majorana particle
that emerge from the Dirac equation and the Majorana condition when
a representation is chosen, it is important to realize that in those
cases where a complex first-order equation for the upper or lower
single component of the entire wave function can be written (for example,
in the Dirac representation), the respective lower or upper single
component is automatically determined by the Majorana condition (depending
on the space-time dimension, this single component can be a two-component
or a one-component wave function). The entire wave function that describes
the Majorana particle can be immediately constructed from these two
components (the upper and the lower components). However, as explained
above, the entire wave function is not absolutely needed to describe
the Majorana particle; in fact, although the upper or lower component
and its respective lower or upper component are not independent wave
functions (i.e., they are not independent of each other), each of
them satisfies its own equation, and either of these two can be considered
as modeling the Majorana particle.

In (3+1) dimensions, there exists an equation for the upper component
and another for the lower component that stand out above the rest
(in this case, these components are two-component wave functions);
these are the ones that arise when the Weyl representation is used.
In fact, each of these equations can also be written in an explicitly
Lorentz covariant form and can describe a specific type of Majorana
particle. These equations have been named the two-component Majorana
equations and tend to the usual Weyl equations when the mass of the
particle and the scalar potential go to zero \cite{RefH,RefI}. Apropos
of the latter result, in (1+1) dimensions and also in the Weyl representation,
we have instead a complex system of coupled equations, i.e., in this
case, we cannot write a first-order equation for any of the components
of the wave function. 

On the other hand, in (3+1) and (1+1) dimensions and in the Majorana
representation, we also have a real system of coupled equations, and
again, no first-order equation for any of the components of the wave
function exists. Finally, the present contribution, beyond clarifying
how the Majorana particle is described (in first quantization), also
attempts to show the different forms of the equations that can arise
when describing it, both in (3+1) and in (1+1) dimensions. We believe
that a detailed discussion on these issues could be useful and quite
pertinent.

The article is organized as follows. In section II, we present the
most basic results that have to do with the relativistic wave equation
commonly used to describe a Majorana particle, namely, the Dirac equation
with a real Lorentz scalar potential. These results are presented
for the cases of (3+1) and (1+1) dimensions. 

In section III, we introduce the charge-conjugation matrix in each
of the representations that we consider in the paper. We use only
three representations, namely, Dirac (or the standard representation),
Weyl (or the chiral or spinor representation) and Majorana. In practice,
these are the most used; the first of these makes it very convenient
to discuss the non-relativistic limit, the second makes immediately
visible the relativistic invariance of the Dirac equation and is very
useful for studying very fast particles, and the third could lead
to real solutions for the Dirac equation with a real Lorentz scalar
potential. Certainly, physics cannot depend on the choice of representation,
although which representation is the best choice depends on the physics.
In this section, the charge-conjugation matrices are obtained from
a good formula that relates the matrices of charge conjugation in
any two representations with the respective similarity matrix that
changes the gamma matrices between these two representations. However,
we specifically use the fact that in the Majorana representation,
the charge-conjugation matrix is the identity matrix; thus, the charge-conjugation
matrix in any representation is a function of the similarity matrix
that takes us from that representation to the Majorana representation.
Again, all these results are presented for the cases of (3+1) and
(1+1) dimensions. 

In section IV, we first present the condition that defines the Majorana
particle, i.e., the Majorana condition. We then present the equations
and systems of equations that come out of the Dirac equation with
a real Lorentz scalar potential and the restriction imposed by the
Majorana condition. Again, we consider the Dirac, Weyl, and Majorana
representations, both in (3+1) and (1+1) dimensions. We also highlight
here the similarities and differences that arise between these equations
in a specific representation but in a different space-time dimension.
In this regard, we note that, in the Weyl representation, there is
a deeper and unexpected difference between these equations. Likewise,
we highlight in this section the procedure that leads us in certain
cases to write the entire wave function from the solution of a single
equation and the Majorana condition (although the solution of this
equation can be sufficient in the description of the Majorana particle).
In this section, we also introduce, for the first time, various results
related to the boundary conditions that can be imposed on the respective
wave function that describes the one-dimensional Majorana particle
in a box, in the Weyl representation.

To complete our study, in section V, we first rederive in detail an
algebraic procedure introduced some time ago by Case to obtain, from
the Dirac equation in (3+1) dimensions and the Majorana condition,
one of the two two-component Majorana equations \cite{RefH}. In fact,
we also obtain the latter two equations after using the Weyl representation
in our results, as expected. Moreover, we write these equations in
distinct ways and compare these results with others commonly presented
in the literature. Then, we also use the Dirac and Majorana representations
in our results. In addition, we also introduce for the first time
an algebraic procedure somewhat analogous to that of Case but this
time in (1+1) dimensions. Then, we repeat the previous program by
using the Weyl, Dirac, and Majorana representations in these new results.
Throughout this section, we re-obtain the most important results presented
in section IV. Finally, in section VI, we write our conclusions. 

\section{Basic results}

\noindent The equation for a Dirac single-particle in (3+1) dimensions,
in a real-valued Lorentz scalar potential $V_{\mathrm{S}}=V_{\mathrm{S}}(x,y,z,t)=V_{\mathrm{S}}(\mathrm{\mathbf{r}},t)$,
\begin{equation}
\left[\,\mathrm{i}\hat{\gamma}^{\mu}\partial_{\mu}-\frac{1}{\hbar c}(V_{\mathrm{S}}+\mathrm{m}c^{2})\hat{1}_{4}\,\right]\Psi=0,
\end{equation}
is satisfied by the (generally) complex Dirac wave function of four
components $\Psi$. The matrix $\hat{1}_{4}$ is the $4$-dimensional
unit matrix. The matrices $\hat{\gamma}^{\mu}=(\hat{\gamma}^{0},\hat{\gamma}^{j})\equiv(\hat{\beta},\hat{\beta}\hat{\alpha}{}_{j})$,
with $\mu=0,j$ and $j=1,2,3$, are the gamma matrices, and the matrices
$\hat{\alpha}{}_{j}$ and $\hat{\beta}$ are the Dirac matrices. The
latter are Hermitian and satisfy the (Clifford) relations $\{\hat{\alpha}{}_{j},\hat{\beta}\}\equiv\hat{\alpha}{}_{j}\hat{\beta}+\hat{\beta}\hat{\alpha}{}_{j}=\hat{0}_{4}$
($\hat{0}_{4}$ is the $4$-dimensional zero matrix), $\{\hat{\alpha}{}_{j},\hat{\alpha}{}_{k}\}=2\delta_{jk}\hat{1}_{4}$
and $\hat{\beta}^{2}=\hat{1}_{4}$ ($\delta_{jk}$ is the Kronecker
delta). Therefore, $\{\hat{\gamma}^{\mu},\hat{\gamma}^{\nu}\}=2g^{\mu\nu}\hat{1}_{4}$,
where $g^{\mu\nu}=\mathrm{diag}(1,-1,-1,-1)$ is the metric tensor,
and $(\hat{\gamma}^{\mu})^{\dagger}=\hat{\gamma}^{0}\hat{\gamma}^{\mu}\hat{\gamma}^{0}$
($\dagger$ denotes the Hermitian conjugate, or the adjoint, of a
matrix and an operator, as usual). The latter two relations imply
that the gamma matrices are unitary, but only $\hat{\gamma}^{0}$
is Hermitian, while $\hat{\gamma}^{j}$ is anti-Hermitian. 

Multiplying Eq. (1) from the left by the operator $\mathrm{i}\hat{\gamma}^{\mu}\partial_{\mu}+\frac{1}{\hbar c}(V_{\mathrm{S}}+\mathrm{m}c^{2})\hat{1}_{4}$
leads to the following second-order equation:
\begin{equation}
\left[\,\hat{1}_{4}\,\partial^{\mu}\partial_{\mu}+\frac{1}{\hbar c}(\partial_{\mu}V_{\mathrm{S}})\,\mathrm{i}\hat{\gamma}^{\mu}+\frac{\left(V_{\mathrm{S}}+\mathrm{m}c^{2}\right)^{2}}{\hbar^{2}c^{2}}\hat{1}_{4}\,\right]\Psi=0.
\end{equation}
Notice that the term containing $\hat{\gamma}^{\mu}$ is not generally
a diagonal matrix, then the components of $\Psi$ mix, In the free
case ($V_{\mathrm{S}}=\mathrm{const}$), all the components satisfy
the same equation, namely, the Klein-Fock-Gordon equation with mass
$\mathrm{m}c^{2}+\mathrm{const}$. Thus, the solutions of the Dirac
equation with a Lorentz scalar potential, i.e., its components, must
comply with a second-order equation. 

The Dirac equation, written in its canonical form, is
\begin{equation}
\left(\mathrm{i}\hbar\hat{1}_{4}\,\frac{\partial}{\partial t}-\hat{\mathrm{H}}\right)\Psi=0,
\end{equation}
where the Hamiltonian operator $\hat{\mathrm{H}}$ is 
\begin{equation}
\hat{\mathrm{H}}=-\mathrm{i}\hbar c\left(\hat{\alpha}{}_{1}\frac{\partial}{\partial x}+\hat{\alpha}{}_{2}\frac{\partial}{\partial y}+\hat{\alpha}{}_{3}\frac{\partial}{\partial z}\right)+(V_{\mathrm{S}}+\mathrm{m}c^{2})\hat{\beta}.
\end{equation}
Eq. (3) is obtained from Eq. (1) by multiplying it by the matrix $\hbar c\hat{\gamma}^{0}=\hbar c\hat{\beta}$
from the left, and using the relations ($\hat{\gamma}^{0})^{2}=\hat{1}_{4}$
and $\hat{\gamma}^{0}\hat{\gamma}^{j}=\hat{\alpha}{}_{j}$. 

Likewise, Eq. (1) is also satisfied by the charge-conjugate wave function
$\Psi_{C}$, but this yields 
\begin{equation}
\hat{S}_{C}\,(-\hat{\gamma}^{\mu})^{*}(\hat{S}_{C})^{-1}=\hat{\gamma}^{\mu}\,,\quad\mathrm{where}\quad\Psi_{C}\equiv\hat{S}_{C}\,\Psi^{*},
\end{equation}
and $\hat{S}_{C}$ is the charge-conjugation matrix (the asterisk
$^{*}$ represents the complex conjugate) \cite{RefJ,RefK}. This
matrix is obviously determined up to a phase factor. As we said before,
the matrices $\hat{\gamma}^{\mu}$ are unitary. More specifically,
this is because $\{\hat{\gamma}^{0},\hat{\gamma}^{j}\}=\hat{0}_{4}$
and $(\hat{\gamma}^{0})^{2}=-(\hat{\gamma}^{j})^{2}=\hat{1}_{4}$
and because $(\hat{\gamma}^{0})^{\dagger}=\hat{\gamma}^{0}\hat{\gamma}^{0}\hat{\gamma}^{0}$
and $(\hat{\gamma}^{j})^{\dagger}=\hat{\gamma}^{0}\hat{\gamma}^{j}\hat{\gamma}^{0}$.
Likewise, the matrices $(-\hat{\gamma}^{\mu})^{*}$ are also unitary.
In effect, $g^{\mu\nu}$ is real; thus, we can write $(-\hat{\gamma}^{\mu})^{*}(-\hat{\gamma}^{\nu})^{*}+(-\hat{\gamma}^{\nu})^{*}(-\hat{\gamma}^{\mu})^{*}=2g^{\mu\nu}\hat{1}_{4}$,
and $((-\hat{\gamma}^{\mu})^{*})^{\dagger}=(-\hat{\gamma}^{0})^{*}(-\hat{\gamma}^{\mu})^{*}(-\hat{\gamma}^{0})^{*}$;
therefore, $((-\hat{\gamma}^{0})^{*})^{\dagger}=((-\hat{\gamma}^{0})^{*})^{-1}$
and $((-\hat{\gamma}^{j})^{*})^{\dagger}=((-\hat{\gamma}^{j})^{*})^{-1}$.
Thus, because the matrices $\hat{\gamma}^{\mu}$ and $(-\hat{\gamma}^{\mu})^{*}$
are linked via the relation on the left side of Eq. (5), the matrix
$\hat{S}_{C}$ can be chosen to be unitary (for more details on this
result, see, for example, Ref. \cite{RefL}, pag. 899). For example,
in the Majorana representation, we have that $\Psi_{C}=\Psi^{*}$,
i.e., $\hat{S}_{C}=\hat{1}_{4}$, and that $\hat{\gamma}^{\mu}=(-\hat{\gamma}^{\mu})^{*}=\mathrm{i}\,\mathrm{\mathrm{Im}}(\hat{\gamma}^{\mu})$
(by virtue of Eq. (5)), i.e., all the entries of the gamma matrices
are purely imaginary. Also, we have that $\mathrm{i}\hat{\gamma}^{\mu}=(\mathrm{i}\hat{\gamma}^{\mu})^{*}=\mathrm{Re}(\mathrm{i}\hat{\gamma}^{\mu})$,
and consequently, the operator acting on $\Psi$ in Eq. (1) is real.
The latter condition implies only that Eq. (1) could have real-valued
solutions. In the same way, Eq. (2) could also have real solutions.

All the equations and relations that we have written so far in (3+1)
dimensions and that are dependent on Greek and Latin indices maintain
their form in (1+1) dimensions. Certainly, these indices are now restricted
to $\mu,\nu,\mathrm{etc}=0,1$, and $j,k,\mathrm{etc}=1$. The Dirac
wave function $\Psi$ now has only two components and satisfies Eqs.
(1), (2) and (3), with $\hat{1}_{4}\rightarrow\hat{1}_{2}$ ($\hat{1}_{2}$
is the $2\times2$ identity matrix) also $V_{\mathrm{S}}=V_{\mathrm{S}}(x,t)$.
The gamma matrices are just $\hat{\gamma}^{0}\equiv\hat{\beta}$ and
$\hat{\gamma}^{1}\equiv\hat{\beta}\hat{\alpha}$, where the (Hermitian)
Dirac matrices, $\hat{\alpha}$ and $\hat{\beta}$, satisfy the relations
$\{\hat{\alpha},\hat{\beta}\}=\hat{0}_{2}$ ($\hat{0}_{2}$ is the
$2$-dimensional zero matrix), $\hat{\alpha}^{2}=\hat{1}_{2}$ and
$\hat{\beta}^{2}=\hat{1}_{2}$. Thus, $\{\hat{\gamma}^{\mu},\hat{\gamma}^{\nu}\}=2g^{\mu\nu}\hat{1}_{2}$,
where $g^{\mu\nu}=\mathrm{diag}(1,-1)$, and $(\hat{\gamma}^{\mu})^{\dagger}=\hat{\gamma}^{0}\hat{\gamma}^{\mu}\hat{\gamma}^{0}$.
As before, the two gamma matrices are unitary, but $\hat{\gamma}^{0}$
is Hermitian, and $\hat{\gamma}^{1}$ is anti-Hermitian. Likewise,
the Hamiltonian operator for the Dirac equation in Eq. (3) is simply
given by
\begin{equation}
\hat{\mathrm{H}}=-\mathrm{i}\hbar c\,\hat{\alpha}\frac{\partial}{\partial x}+(V_{\mathrm{S}}+\mathrm{m}c^{2})\hat{\beta}.
\end{equation}

\section{Charge-conjugation in the Dirac, Weyl, and Majorana representations}

\noindent As is well known, in choosing a representation one is choosing
a set of Dirac and gamma matrices that satisfies a Clifford relation
(i.e., they form a Clifford algebra). As was demonstrated, for instance,
in Ref. \cite{RefF}, if one has written the charge-conjugation matrix
in a representation, let's say $\hat{S}_{C}$, then one can write
it in any other representation, let's say $\hat{S}_{C}'$, via the
following relation:
\begin{equation}
\hat{S}_{C}'=\hat{S}\,\hat{S}_{C}\,(\hat{S}^{*})^{-1},
\end{equation}
where $\hat{S}$ is precisely the unitary similarity matrix that allows
us to pass the unitary gamma matrices between these two representations,
i.e., $\hat{\gamma}^{\mu}\,'=\hat{S}\,\hat{\gamma}^{\mu}\hat{S}^{-1}$.
The result in Eq. (7) is simply due to the fact that the wave functions
$\Psi$ and $\Psi_{C}$ are transformed under $\hat{S}$ as $\Psi'=\hat{S}\,\Psi$
and $\Psi'_{C}=\hat{S}\,\Psi_{C}$, but in each representation we
also have that $\Psi_{C}\equiv\hat{S}_{C}\,\Psi^{*}$ and $\Psi'_{C}\equiv\hat{S}_{C}'\,(\Psi')^{*}$.
Obviously, if we change the phase factor of the matrix $\hat{S}_{C}$,
the matrix $\hat{S}_{C}'$ that is obtained from Eq. (7) changes in
a factor that is also a phase. However, all the matrices involved
in Eq. (7) are always determined up to an arbitrary phase factor.
If we particularize the formula in Eq. (7) to the case in which $\hat{S}_{C}$
is written in an arbitrary representation and $\hat{S}_{C}'$ is written
in the Majorana representation, i.e., $\hat{S}_{C}'=\hat{1}_{4}$,
in (3+1) dimensions or $\hat{S}_{C}'=\hat{1}_{2}$, in (1+1) dimensions,
one obtains the result
\begin{equation}
\hat{S}_{C}=\hat{S}^{\dagger}\hat{S}^{*},
\end{equation}
where $\hat{S}$ is the unitary matrix that takes us from that arbitrary
representation to the Majorana representation. From Eq. (8), and because
$\hat{S}_{C}$ is a unitary matrix, we obtain the result $(\hat{S}_{C})^{-1}=(\hat{S}_{C})^{*}$.
The latter can also be obtained just by requiring that $(\Psi_{C})_{C}=\Psi$.

The results pertinent to those representations usually identified
as Dirac, Weyl, and Majorana in (3+1) dimensions are given in Table
1. The latter also shows the charge-conjugation matrix $\hat{S}_{C}$
in each representation derived from Eq. (8), the respective matrices
$\hat{S}$ being the following:
\begin{equation}
\hat{S}=\frac{1}{\sqrt{2}}\left(\hat{\sigma}_{x}\otimes\hat{\sigma}_{y}\,+\,\hat{\sigma}_{z}\otimes\hat{1}_{2}\right),
\end{equation}
which permits us to pass from the Dirac representation to the Majorana
representation, and 
\begin{equation}
\hat{S}=\frac{1}{2}\left(\hat{\sigma}_{x}\otimes\hat{\sigma}_{y}\,+\,\hat{\sigma}_{z}\otimes\hat{\sigma}_{y}\,+\,\hat{\sigma}_{z}\otimes\hat{1}_{2}\,-\,\hat{\sigma}_{x}\otimes\hat{1}_{2}\right),
\end{equation}
which permits us to pass from the Weyl representation to the Majorana
representation. Obviously, the matrix $\hat{S}=\hat{1}_{4}=\hat{1}_{2}\otimes\hat{1}_{2}$
permits us to pass from the Majorana representation to the Majorana
representation itself. Note that in (3+1) dimensions the charge-conjugation
matrix in the Dirac representation is equal to the charge-conjugation
matrix in the Weyl representation (up to a phase factor). For the
sake of completeness, the matrix $\hat{S}$ that allows us to pass
precisely from the Dirac representation to that of Weyl is also given
here: 
\begin{equation}
\hat{S}=\frac{1}{\sqrt{2}}\left(\hat{1}_{2}\otimes\hat{1}_{2}\,+\,\mathrm{i}\hat{\sigma}_{y}\otimes\hat{1}_{2}\right).
\end{equation}
This matrix links the matrices $\hat{S}_{C}$ (in the Dirac representation)
and $\hat{S}_{C}'$ (in the Weyl representation) also via Eq. (7). 

In reading Tables 1, 1.1 and 1.2, the following definitions should
be considered: $\hat{\boldsymbol{\alpha}}\equiv(\hat{\alpha}_{1},\hat{\alpha}_{2},\hat{\alpha}_{3})$,
$\hat{\boldsymbol{\gamma}}\equiv(\hat{\gamma}^{1},\hat{\gamma}^{2},\hat{\gamma}^{3})$,
and the usual Pauli matrices are $\hat{\boldsymbol{\sigma}}\equiv(\hat{\sigma}_{x},\hat{\sigma}_{y},\hat{\sigma}_{z})$.
$\otimes$ indicates the Kronecker product of matrices 
\begin{equation}
\hat{A}\otimes\hat{B}\equiv\left[\begin{array}{ccc}
a_{11}\hat{B} & \cdots & a_{1n}\hat{B}\\
\vdots & \ddots & \vdots\\
a_{m1}\hat{B} & \cdots & a_{mn}\hat{B}
\end{array}\right],
\end{equation}
which satisfies the following properties: (i) $(\hat{A}\otimes\hat{B})(\hat{C}\otimes\hat{D})=(\hat{A}\hat{C}\otimes\hat{B}\hat{D})$,
(ii) $(\hat{A}\otimes\hat{B})^{*}=\hat{A}^{*}\otimes\hat{B}^{*}$,
and (iii) $(\hat{A}\otimes\hat{B})^{\dagger}=\hat{A}^{\dagger}\otimes\hat{B}^{\dagger}$
(for example, see Ref. \cite{RefM}). Note that here we have $\hat{S}_{C}=-\hat{\gamma}^{2}$
in both the Dirac and Weyl representations. However, when considering
these two representations, it is also common to write $\hat{S}_{C}=+\hat{\gamma}^{2}$,
and in particle physics, it is more common to set $\hat{S}_{C}=+\mathrm{i}\hat{\gamma}^{2}$
and $\hat{S}_{C}=-\mathrm{i}\hat{\gamma}^{2}$. 

In the same way, the results pertinent to those representations commonly
considered as representations of Dirac, Weyl and Majorana in (1+1)
dimensions are given in Table 2. The latter also shows the charge-conjugation
matrix $\hat{S}_{C}$ in each representation calculated from Eq. (8).
The respective matrices $\hat{S}$ are the following:
\begin{equation}
\hat{S}=\frac{1}{\sqrt{2}}\left(\hat{1}_{2}+\mathrm{i}\hat{\sigma}_{x}\right),
\end{equation}
which permits us to pass from the Dirac representation to the Majorana
representation, and
\begin{equation}
\hat{S}=\frac{1}{2}\left(\mathrm{i}\hat{1}_{2}+\hat{\sigma}_{x}+\hat{\sigma}_{y}+\hat{\sigma}_{z}\right),
\end{equation}
which permits us to pass from the Weyl representation to the Majorana
representation. Note that in (1+1) dimensions, the charge-conjugation
matrix in the Dirac representation is not equal to the charge-conjugation
matrix in the Weyl representation. For the sake of completeness, the
matrix $\hat{S}$, which allows us to pass precisely from the Dirac
representation to that of Weyl, is also given here:
\begin{equation}
\hat{S}=\frac{1}{\sqrt{2}}\left(\hat{\sigma}_{x}+\hat{\sigma}_{z}\right).
\end{equation}
This matrix links the matrices $\hat{S}_{C}$ (in the Dirac representation)
and $\hat{S}_{C}'$ (in the Weyl representation) also via Eq. (7).

\section{Equations for the Majorana single-particle I}

\noindent The condition that defines a Majorana particle, called the
Majorana condition, is given by 
\begin{equation}
\Psi=\Psi_{C}=\hat{S}_{C}\,\Psi^{*}.
\end{equation}
In general, the equation that describes this single particle is the
Dirac equation (Eq. (1)) together with the latter condition. Apropos
of this, it is important to note that the wave functions $\Psi\equiv\left[\,\mathrm{top}\;\,\mathrm{bottom}\,\right]^{\mathrm{T}}$
and $\Psi_{C}\equiv\left[\,\mathrm{top}_{C}\;\,\mathrm{bottom}_{C}\,\right]^{\mathrm{T}}$
(where the words ``$\mathrm{top}$'' and ``$\mathrm{bottom}$''
indicate the upper and lower components, respectively, of the respective
wave function) are similarly transformed under proper Lorentz transformations
($^{\mathrm{T}}$ represents the transpose of a matrix). Thus, the
upper components of these two wave functions, as well as the lower
components, are similarly transformed. Obviously, this is true in
any representation and has nothing to do with the Majorana condition.
If in addition, the Majorana condition in Eq. (16) is verified, then
the upper components of $\Psi$ and $\Psi_{C}$, as well as their
lower components, are equal. In passing, the Majorana condition is
sometimes written as $\Psi=\omega\Psi_{C}$, where $\omega$ is an
arbitrary unobservable phase factor, and it is still a Lorentz covariant
condition \cite{RefC}, as expected. Below, we present the equations
or systems of equations for the Majorana particle in the Dirac, Weyl
and Majorana representations both in (3+1) and (1+1) dimensions. We
make full use of Tables 1 and 2.

\subsection{Dirac representation}

\noindent \textbf{In (3+1) dimensions.} We write the four-component
Dirac wave function (or Dirac spinor) $\Psi$ in the form
\begin{equation}
\Psi\equiv\left[\begin{array}{c}
\varphi\\
\chi
\end{array}\right],
\end{equation}
where the upper two-component wave function can be written as $\varphi\equiv\left[\,\varphi_{1}\;\varphi_{2}\,\right]^{\mathrm{T}}$
and the lower one as $\chi\equiv\left[\,\chi_{1}\;\chi_{2}\,\right]^{\mathrm{T}}$.
In (3+1) dimensions, a two-component wave function such as $\Psi$
is also called bispinor. The Dirac equation (Eq. (3)) takes the form
\begin{equation}
\mathrm{i}\hbar\frac{\partial}{\partial t}\left[\begin{array}{c}
\varphi\\
\chi
\end{array}\right]=\hat{\mathrm{H}}\left[\begin{array}{c}
\varphi\\
\chi
\end{array}\right]=\left[\begin{array}{cc}
(V_{\mathrm{S}}+\mathrm{m}c^{2})\hat{1}_{2} & -\mathrm{i}\hbar c\,\hat{\boldsymbol{\sigma}}\cdot\nabla\\
-\mathrm{i}\hbar c\,\hat{\boldsymbol{\sigma}}\cdot\nabla & -(V_{\mathrm{S}}+\mathrm{m}c^{2})\hat{1}_{2}
\end{array}\right]\left[\begin{array}{c}
\varphi\\
\chi
\end{array}\right].
\end{equation}
The Majorana condition (Eq. (16)) imposed upon the Dirac wave function
imposes the following relation among the components of $\Psi$:
\begin{equation}
\chi=\hat{\sigma}_{y}\,\varphi^{*}\equiv\chi_{C}\quad(\,\Leftrightarrow\,\varphi=-\hat{\sigma}_{y}\,\chi^{*}\equiv\varphi_{C}).
\end{equation}
Substituting the latter $\chi$ into Eq. (18), we are left with an
equation for the two-component wave function $\varphi$, namely, 
\begin{equation}
\mathrm{i}\hbar\hat{1}_{2}\,\frac{\partial}{\partial t}\varphi=-\mathrm{i}\hbar c\,\hat{\boldsymbol{\sigma}}\cdot\nabla\left(\hat{\sigma}_{y}\,\varphi^{*}\right)+(V_{\mathrm{S}}+\mathrm{m}c^{2})\hat{1}_{2}\varphi.
\end{equation}
Certainly, by making the latter replacement, two equations arise:
one is Eq. (20), and the other is an equation that can also be obtained
from Eq. (20) by making the following substitutions: $\varphi\rightarrow\hat{\sigma}_{y}\,\varphi^{*}$,
$\hat{\sigma}_{y}\,\varphi^{*}\rightarrow\varphi$, and $V_{\mathrm{S}}+\mathrm{m}c^{2}\rightarrow-(V_{\mathrm{S}}+\mathrm{m}c^{2})$.
Then, it can be algebraically shown that the latter equation and Eq.
(20) are equivalent. Alternatively, if we substitute $\varphi$ (from
Eq. (19)) into Eq. (18), we obtain the following equation for the
two-component wave function $\chi$:
\begin{equation}
\mathrm{i}\hbar\hat{1}_{2}\,\frac{\partial}{\partial t}\chi=-\mathrm{i}\hbar c\,\hat{\boldsymbol{\sigma}}\cdot\nabla\left(-\hat{\sigma}_{y}\,\chi^{*}\right)-(V_{\mathrm{S}}+\mathrm{m}c^{2})\hat{1}_{2}\chi.
\end{equation}
Again, by making the latter replacement, two equations arise: one
is Eq. (21), and the other is an equation that can also be obtained
from Eq. (21) by making the following replacements: $\chi\rightarrow-\hat{\sigma}_{y}\,\chi^{*}$,
$-\hat{\sigma}_{y}\,\chi^{*}\rightarrow\chi$, and $-(V_{\mathrm{S}}+\mathrm{m}c^{2})\rightarrow V_{\mathrm{S}}+\mathrm{m}c^{2}$.
Again, it can be algebraically shown that the latter equation and
Eq. (21) are absolutely equivalent. Clearly, if we assume that the
wave function that describes the Majorana particle has four components,
it is sufficient to solve at least one of the two last (decoupled)
two-component equations, namely, Eqs. (20) and (21). This is because
$\varphi$ and $\chi$ are algebraically related by Eq. (19). Thus,
Eq. (20) (or Eq. (21)) alone can be considered as a two-component
equation that models the Majorana particle in (3+1) dimensions.

In Ref. \cite{RefN}, an equation analogous to Eq. (20), with $V_{\mathrm{S}}=0$,
was recently related to a nonlocal Schr\"odinger-type equation. Interestingly
enough, the latter equation does not suffer from some of the problems
that typically adversely affect relativistic single-particle equations.
Incidentally, the authors in Ref. \cite{RefN} used the same matrices
corresponding to the Dirac representation shown in Table 1 but used
a slightly different charge-conjugation matrix, namely, $\hat{S}_{C}=+\mathrm{i}\hat{\sigma}_{y}\otimes\hat{\sigma}_{y}$.
Thus, the two-component equation used in that paper is precisely Eq.
(20) with the following replacement: $\hat{\sigma}_{y}\,\varphi^{*}\rightarrow-\hat{\sigma}_{y}\,\varphi^{*}$.
Likewise, in the same reference, two one-parameter families of confining
boundary conditions were obtained for Majorana fermions restricted
to a three-dimensional finite spatial domain.

\noindent \textcompwordmark{}

\noindent \textbf{In (1+1) dimensions.} We write the two-component
Dirac wave function $\Psi$ in the form given in Eq. (17), but in
this case, $\varphi$ and $\chi$ are simply functions of a single
component. The Dirac equation (Eq. (3)) with the Hamiltonian in Eq.
(6) takes the form
\begin{equation}
\mathrm{i}\hbar\frac{\partial}{\partial t}\left[\begin{array}{c}
\varphi\\
\chi
\end{array}\right]=\hat{\mathrm{H}}\left[\begin{array}{c}
\varphi\\
\chi
\end{array}\right]=\left[\begin{array}{cc}
V_{\mathrm{S}}+\mathrm{m}c^{2} & -\mathrm{i}\hbar c\frac{\partial}{\partial x}\\
-\mathrm{i}\hbar c\frac{\partial}{\partial x} & -(V_{\mathrm{S}}+\mathrm{m}c^{2})
\end{array}\right]\left[\begin{array}{c}
\varphi\\
\chi
\end{array}\right].
\end{equation}
The Majorana condition (Eq. (16)) imposed upon the Dirac wave function
imposes the following relation between the two components of $\Psi$:
\begin{equation}
\chi=-\mathrm{i}\varphi^{*}\equiv\chi_{C}\quad(\,\Leftrightarrow\,\varphi=-\mathrm{i}\chi^{*}\equiv\varphi_{C}).
\end{equation}
Substituting the latter $\chi$ into Eq. (22), we are left with an
equation for the one-component wave function $\varphi$, namely,
\begin{equation}
\mathrm{i}\hbar\frac{\partial}{\partial t}\varphi=-\mathrm{i}\hbar c\frac{\partial}{\partial x}(-\mathrm{i}\varphi^{*})+(V_{\mathrm{S}}+\mathrm{m}c^{2})\varphi
\end{equation}
(the other equation that results after making the previous substitution
in Eq. (22) is essentially the complex conjugate equation of Eq. (24)).
Different from how it is in (3+1) dimensions, the equation for the
lower component $\chi$ is simply equal to the equation for the upper
component (Eq. (24)) but with the following replacement: $V_{\mathrm{S}}+\mathrm{m}c^{2}\rightarrow-(V_{\mathrm{S}}+\mathrm{m}c^{2})$.
In any case, it is sufficient to solve at least one of these one-component
equations because $\varphi$ and $\chi$ are algebraically linked
via Eq. (23). Thus, for example, it can be said that Eq. (24) (or
the equation for $\chi$) alone models the Majorana particle in (1+1)
dimensions \cite{RefF}.

Again, in Ref. \cite{RefN}, an equation analogous to Eq. (24), with
$V_{\mathrm{S}}=0$, was related to a relativistic Schr\"odinger-type
equation that has a consistent quantum mechanical single-particle
interpretation (for example, it does not have negative energy states).
In that reference, the authors used the same Dirac representation
as we do, but the matrix $\hat{S}_{C}=+\mathrm{i}\hat{\sigma}_{x}$
was used instead; thus, the equation for the Majorana particle used
therein is precisely Eq. (24) with the following replacement: $-\mathrm{i}\varphi^{*}\rightarrow+\mathrm{i}\varphi^{*}$.
On the other hand, the only four boundary conditions that $\varphi$
can support when the one-dimensional Majorana particle is inside an
impenetrable box (we call them confining boundary conditions) were
also encountered in Ref. \cite{RefN}. Likewise, these conditions
were found in Ref. \cite{RefF}, but it was shown in the latter reference
that these are just the conditions that can arise mathematically from
the general linear boundary condition used in the MIT bag model for
a hadronic structure in (1+1) dimensions (certainly, the latter four
boundary conditions are also subject to the Majorana condition). Specifically,
for a box of size $L$ with ends, for example, at $x=0$ and $x=L$,
the four confining boundary conditions can be written in the form
$f(0,t)=g(L,t)=0$, where $f$ and $g$ are the functions $\mathrm{Im}(\varphi)$
and $\mathrm{Re}(\varphi)$. Clearly, two of these boundary conditions
are just the Dirichlet boundary condition imposed upon $\mathrm{Im}(\varphi)$
and $\mathrm{Re}(\varphi)$ at the ends of the box. The latter is
a nice result because the entire two-component Dirac wave function
does not support this type of boundary condition at the walls of the
box \cite{RefO}. In addition, two one-parameter families of non-confining
boundary conditions, i.e., infinite non-confining boundary conditions
(we call them non-confining because they do not cancel the probability
current density at the ends of the box), were found in Ref. \cite{RefF}.
It is even possible (by taking some convenient limits) that these
two families also include the four confining boundary conditions.
Consequently, these two families actually make up the most general
set of boundary conditions for the one-dimensional Majorana particle
in a box; see Eq. (93) in Ref. \cite{RefF}. In detail, we write below,
for the first time, these two families of boundary conditions but
in the Weyl representation.

Clearly, in the Dirac representation, the procedure for finding single
equations for the Majorana particle is similar in (3+1) and (1+1)
dimensions. However, this representation is not so commonly used to
write the equation for the Majorana particle, neither in (3+1) nor
in (1+1) dimensions; rather, Weyl's representation is used (at least
in (3+1) dimensions).

\subsection{Weyl representation}

\noindent \textbf{In (3+1) dimensions.} We write the four-component
Dirac wave function (or spinor) $\Psi$ as follows:
\begin{equation}
\Psi\equiv\left[\begin{array}{c}
\varphi_{1}\\
\varphi_{2}
\end{array}\right],
\end{equation}
where the upper (lower) two-component wave function can be written
as $\varphi_{1}\equiv\left[\,\xi_{1}\;\xi_{2}\,\right]^{\mathrm{T}}$
($\varphi_{2}\equiv\left[\,\xi_{3}\;\xi_{4}\,\right]^{\mathrm{T}}$).
The Dirac equation (Eq. (3)) takes the form 
\begin{equation}
\mathrm{i}\hbar\frac{\partial}{\partial t}\left[\begin{array}{c}
\varphi_{1}\\
\varphi_{2}
\end{array}\right]=\hat{\mathrm{H}}\left[\begin{array}{c}
\varphi_{1}\\
\varphi_{2}
\end{array}\right]=\left[\begin{array}{cc}
-\mathrm{i}\hbar c\,\hat{\boldsymbol{\sigma}}\cdot\nabla & -(V_{\mathrm{S}}+\mathrm{m}c^{2})\hat{1}_{2}\\
-(V_{\mathrm{S}}+\mathrm{m}c^{2})\hat{1}_{2} & +\mathrm{i}\hbar c\,\hat{\boldsymbol{\sigma}}\cdot\nabla
\end{array}\right]\left[\begin{array}{c}
\varphi_{1}\\
\varphi_{2}
\end{array}\right].
\end{equation}
The Majorana condition (Eq. (16)) imposed upon $\Psi$ leads us to
the following relation among its components:
\begin{equation}
\varphi_{2}=\hat{\sigma}_{y}\,\varphi_{1}^{*}\equiv(\varphi_{2})_{C}\quad(\,\Leftrightarrow\,\varphi_{1}=-\hat{\sigma}_{y}\,\varphi_{2}^{*}\equiv(\varphi_{1})_{C}\,).
\end{equation}
Substituting the latter $\varphi_{2}$ into Eq. (26), we are left
with an equation for the two-component wave function $\varphi_{1}$,
namely,
\begin{equation}
\mathrm{i}\hbar\hat{1}_{2}\,\frac{\partial}{\partial t}\varphi_{1}=-\mathrm{i}\hbar c\,\hat{\boldsymbol{\sigma}}\cdot\nabla\varphi_{1}-(V_{\mathrm{S}}+\mathrm{m}c^{2})\,\hat{\sigma}_{y}\,\varphi_{1}^{*}.
\end{equation}
Instead, if we substitute $\varphi_{1}$ (from Eq. (27)) into Eq.
(26), we obtain the following equation for the two-component wave
function $\varphi_{2}$:
\begin{equation}
\mathrm{i}\hbar\hat{1}_{2}\,\frac{\partial}{\partial t}\varphi_{2}=+\mathrm{i}\hbar c\,\hat{\boldsymbol{\sigma}}\cdot\nabla\varphi_{2}+(V_{\mathrm{S}}+\mathrm{m}c^{2})\,\hat{\sigma}_{y}\,\varphi_{2}^{*}.
\end{equation}
Again, to obtain the wave function $\Psi$ (Eq. (25)), it is sufficient
to solve first Eq. (28) (Eq. (29)) to obtain $\varphi_{1}$ ($\varphi_{2}$)
and then obtain $\varphi_{2}$ ($\varphi_{1}$) by using the Majorana
condition (Eq. (27)). Note that the substitution that gave us Eq.
(28) for $\varphi_{1}$ also generates another equation, namely, Eq.
(29) for $\hat{\sigma}_{y}\,\varphi_{1}^{*}$ (these two equations
are algebraically equivalent). Likewise, the substitution that gave
us Eq. (29) for $\varphi_{2}$ also generates another equation, namely,
Eq. (28) for $-\hat{\sigma}_{y}\,\varphi_{2}^{*}$ (again, both equations
are equivalent). Thus, the wave function $\varphi_{1}=\varphi_{1}(\mathrm{\mathbf{r}},t)$
satisfies Eq. (28), but unexpectedly, $\mathrm{i}\hat{\sigma}_{y}\,\varphi_{1}^{*}(-\mathrm{\mathbf{r}},t)$
also satisfies Eq. (28) (provided that the relation $V_{\mathrm{S}}(\mathrm{\mathbf{r}},t)=V_{\mathrm{S}}(-\mathrm{\mathbf{r}},t)$
is fulfilled). Similarly, the wave function $\varphi_{2}=\varphi_{2}(\mathrm{\mathbf{r}},t)$
satisfies Eq. (29), but $-\mathrm{i}\hat{\sigma}_{y}\,\varphi_{2}^{*}(-\mathrm{\mathbf{r}},t)$
also satisfies Eq. (29) (and again, the scalar potential must be an
even function in $\mathrm{\mathbf{r}}$).

Making $\mathrm{m}c^{2}=V_{\mathrm{S}}=0$ in Eq. (26), one obtains
two (decoupled) equations, namely,
\begin{equation}
\mathrm{i}\hbar\hat{1}_{2}\,\frac{\partial}{\partial t}\varphi_{1}=-\mathrm{i}\hbar c\,\hat{\boldsymbol{\sigma}}\cdot\nabla\varphi_{1}\,,\quad\mathrm{i}\hbar\hat{1}_{2}\,\frac{\partial}{\partial t}\varphi_{2}=+\mathrm{i}\hbar c\,\hat{\boldsymbol{\sigma}}\cdot\nabla\varphi_{2}.
\end{equation}
These are the well-known Weyl's equations. For instance, the first
of these two-component equations can be assigned to the (right-handed,
or right-helical) massless antineutrino, while the second one can
be assigned to the (left-handed, or left-helical) massless neutrino
(even though it is possible that only one of these two equations is
sufficient for the description of a massless fermion, in which case
one is led to the so-called Weyl theory \cite{RefK,RefP}). On the
other hand, making $\mathrm{m}c^{2}=V_{\mathrm{S}}=0$ in Eqs. (28)
and (29) one obtains two equations (in fact, the same equations given
in Eq. (30)), but this time, $\varphi_{1}$ and $\varphi_{2}$ are
related by the Majorana condition in Eq. (27). In fact, the four-component
Majorana wave functions corresponding to the two-component wave functions
$\varphi_{1}$ and $\varphi_{2}$ are given by 
\begin{equation}
\Psi=\left[\begin{array}{c}
\varphi_{1}\\
\hat{\sigma}_{y}\,\varphi_{1}^{*}
\end{array}\right](=\Psi_{C})\,,\quad\mathrm{and}\quad\Psi=\left[\begin{array}{c}
-\hat{\sigma}_{y}\,\varphi_{2}^{*}\\
\varphi_{2}
\end{array}\right](=\Psi_{C}),
\end{equation}
respectively. Meanwhile, the four-component Weyl wave functions corresponding
to the two-component wave functions $\varphi_{1}$ and $\varphi_{2}$
are given by
\begin{equation}
\Psi=\left[\begin{array}{c}
\varphi_{1}\\
0
\end{array}\right]\,,\quad\mathrm{and}\quad\Psi=\left[\begin{array}{c}
0\\
\varphi_{2}
\end{array}\right],
\end{equation}
respectively. 

As can be seen in Table 1, we have considered the matrices $\hat{\boldsymbol{\alpha}}=+\hat{\sigma}_{z}\otimes\hat{\boldsymbol{\sigma}}$
and $\hat{\beta}=-\hat{\sigma}_{x}\otimes\hat{1}_{2}$ as the Weyl
representation; however, in some books and articles, the matrices
$\hat{\boldsymbol{\alpha}}'=+\hat{\sigma}_{z}\otimes\hat{\boldsymbol{\sigma}}$
and $\hat{\beta}'=+\hat{\sigma}_{x}\otimes\hat{1}_{2}$ are also used
as a Weyl representation (for example, in Refs. \cite{RefI,RefQ}).
Even the matrices $\hat{\boldsymbol{\alpha}}'=-\hat{\sigma}_{z}\otimes\hat{\boldsymbol{\sigma}}$
and $\hat{\beta}'=+\hat{\sigma}_{x}\otimes\hat{1}_{2}$ have been
considered as a Weyl representation in other publications (for example,
in Ref. \cite{RefC}). As usual, the former and the latter two representations
are related by $\hat{\boldsymbol{\alpha}}'=\hat{S}\,\hat{\boldsymbol{\alpha}}\,\hat{S}^{\dagger}$
and $\hat{\beta}'=\hat{S}\,\hat{\beta}\,\hat{S}^{\dagger}$ as well
as $\left[\,\varphi_{1}'\;\varphi_{2}'\,\right]^{\mathrm{T}}=\hat{S}\left[\,\varphi_{1}\;\varphi_{2}\,\right]^{\mathrm{T}}$,
but we must use $\hat{S}=\hat{\sigma}_{z}\otimes\hat{1}_{2}=\hat{S}^{\dagger}$
to relate the first and the second pair of Dirac matrices and $\hat{S}=\hat{\sigma}_{y}\otimes\hat{1}_{2}=\hat{S}^{\dagger}$
to relate the first and the third pair.

In (3+1) dimensions, the Weyl representation is definitely the most
used. As we will show in section V, the two-component Eqs. (28) and
(29) can also be written explicitly in covariant form, and each of
them can describe a Majorana particle. 

\noindent \textcompwordmark{}

\noindent \textbf{In (1+1) dimensions.} We write the two-component
Dirac wave function $\Psi$ in the form given in Eq. (25), but in
this case, $\varphi_{1}$ and $\varphi_{2}$ are wave functions of
a single component. The Dirac equation (Eq. (3)) takes the form
\begin{equation}
\mathrm{i}\hbar\frac{\partial}{\partial t}\left[\begin{array}{c}
\varphi_{1}\\
\varphi_{2}
\end{array}\right]=\hat{\mathrm{H}}\left[\begin{array}{c}
\varphi_{1}\\
\varphi_{2}
\end{array}\right]=\left[\begin{array}{cc}
-\mathrm{i}\hbar c\frac{\partial}{\partial x} & V_{\mathrm{S}}+\mathrm{m}c^{2}\\
V_{\mathrm{S}}+\mathrm{m}c^{2} & +\mathrm{i}\hbar c\frac{\partial}{\partial x}
\end{array}\right]\left[\begin{array}{c}
\varphi_{1}\\
\varphi_{2}
\end{array}\right].
\end{equation}
The Majorana condition (Eq. (16)) imposed upon $\Psi$ gives us the
following relations:
\begin{equation}
\varphi_{1}=-\mathrm{i}\varphi_{1}^{*}\equiv(\varphi_{1})_{C}\:\,\mathrm{and}\:\,\varphi_{2}=+\mathrm{i}\varphi_{2}^{*}\equiv(\varphi_{2})_{C}.
\end{equation}
Obviously, these relations do not allow us to write a one-component
first-order equation for the Majorana particle (and from Eq. (2),
neither can a standard one-component second-order equation be written).
That is, unlike what happens in (3+1) dimensions, the equation that
describes the Majorana particle in (1+1) dimensions is a complex system
of coupled equations, i.e., Eq. (33) with the restriction given in
Eq. (34). 

In this representation, we can also write the most general set of
boundary conditions for the one-dimensional Majorana particle inside
a box with ends at $x=0$ and $x=L$. This set consists of two one-parameter
families of boundary conditions. In fact, using the results given
in Eqs. (67) and (68) of Ref. \cite{RefF} (written in the Majorana
representation) and the fact that the two-component wave functions
in the Weyl and Majorana representations verify the relation $\left[\,\phi_{1}\;\phi_{2}\,\right]^{\mathrm{T}}=\hat{S}\left[\,\varphi_{1}\;\varphi_{2}\,\right]^{\mathrm{T}}$,
where the matrix $\hat{S}$ is given in Eq. (14), we obtain, respectively
(we exclude the variable $t$ in the boundary conditions hereinafter),
\begin{equation}
\left[\begin{array}{c}
\varphi_{1}(L)\\
\varphi_{2}(L)
\end{array}\right]=\frac{1}{m_{2}}\left[\begin{array}{cc}
-1 & -\mathrm{i}m_{0}\\
-\mathrm{i}m_{0} & +1
\end{array}\right]\left[\begin{array}{c}
\varphi_{1}(0)\\
\varphi_{2}(0)
\end{array}\right],
\end{equation}
where $(m_{0})^{2}+(m_{2})^{2}=1$, and
\begin{equation}
\left[\begin{array}{c}
\varphi_{1}(L)\\
\varphi_{2}(L)
\end{array}\right]=\frac{1}{m_{1}}\left[\begin{array}{cc}
+1 & -\mathrm{i}m_{3}\\
+\mathrm{i}m_{3} & +1
\end{array}\right]\left[\begin{array}{c}
\varphi_{1}(0)\\
\varphi_{2}(0)
\end{array}\right],
\end{equation}
where $(m_{1})^{2}+(m_{3})^{2}=1$. Note that the $2\times2$ matrix
in (35) is equal to its own inverse and that the inverse matrix of
the $2\times2$ matrix in (36) is obtained from the latter by making
the substitution $m_{3}\rightarrow-m_{3}$. We obtain two boundary
conditions for an impenetrable box (i.e., two confining boundary conditions)
from Eq. (35) and its inverse by making $m_{2}\rightarrow0$, namely,
\begin{equation}
\varphi_{1}(L)=-\mathrm{i}\varphi_{2}(L)\,,\quad\varphi_{1}(0)=-\mathrm{i}\varphi_{2}(0),
\end{equation}
with $m_{0}=1$, and
\begin{equation}
\varphi_{1}(L)=+\mathrm{i}\varphi_{2}(L)\,,\quad\varphi_{1}(0)=+\mathrm{i}\varphi_{2}(0),
\end{equation}
with $m_{0}=-1$. Likewise, we obtain two other confining boundary
conditions from Eq. (36) and its inverse by making $m_{1}\rightarrow0$,
namely,
\begin{equation}
\varphi_{1}(L)=-\mathrm{i}\varphi_{2}(L)\,,\quad\varphi_{1}(0)=+\mathrm{i}\varphi_{2}(0),
\end{equation}
with $m_{3}=1$, and
\begin{equation}
\varphi_{1}(L)=+\mathrm{i}\varphi_{2}(L)\,,\quad\varphi_{1}(0)=-\mathrm{i}\varphi_{2}(0),
\end{equation}
with $m_{3}=-1$. Note that the wave function $\left[\,\varphi_{1}\;\varphi_{2}\,\right]^{\mathrm{T}}$
can satisfy any of the boundary conditions included in Eqs. (35) and
(36), but then the wave function $\left[\,-\mathrm{i}\varphi_{1}^{*}\;+\mathrm{i}\varphi_{2}^{*}\,\right]^{\mathrm{T}}$
also automatically satisfies this boundary condition. This is due
to the Majorana condition. Because in this case the Majorana condition
is a pair of independent relations, the boundary conditions are presented
in terms of the two components of the wave function.

\subsection{Majorana representation}

\noindent \textbf{In (3+1) dimensions.} The four-component Dirac wave
function (or spinor) $\Psi$ can be written as 
\begin{equation}
\Psi\equiv\left[\begin{array}{c}
\phi_{1}\\
\phi_{2}
\end{array}\right],
\end{equation}
where the upper (lower) two-component wave function could be written
as $\phi_{1}\equiv\left[\,\zeta_{1}\;\zeta_{2}\,\right]^{\mathrm{T}}$
($\phi_{2}\equiv\left[\,\zeta_{3}\;\zeta_{4}\,\right]^{\mathrm{T}}$).
The Dirac equation (Eq. (3)) takes the form
\[
\mathrm{i}\hbar\frac{\partial}{\partial t}\left[\begin{array}{c}
\phi_{1}\\
\phi_{2}
\end{array}\right]=\hat{\mathrm{H}}\left[\begin{array}{c}
\phi_{1}\\
\phi_{2}
\end{array}\right]
\]
\begin{equation}
=\left[\begin{array}{cc}
-\mathrm{i}\hbar c\hat{1}_{2}\frac{\partial}{\partial y} & +\mathrm{i}\hbar c\left(\hat{\sigma}_{x}\frac{\partial}{\partial x}+\hat{\sigma}_{z}\frac{\partial}{\partial z}\right)+(V_{\mathrm{S}}+\mathrm{m}c^{2})\hat{\sigma}_{y}\\
+\mathrm{i}\hbar c\left(\hat{\sigma}_{x}\frac{\partial}{\partial x}+\hat{\sigma}_{z}\frac{\partial}{\partial z}\right)+(V_{\mathrm{S}}+\mathrm{m}c^{2})\hat{\sigma}_{y} & +\mathrm{i}\hbar c\hat{1}_{2}\frac{\partial}{\partial y}
\end{array}\right]\left[\begin{array}{c}
\phi_{1}\\
\phi_{2}
\end{array}\right].
\end{equation}
Clearly, Eq. (42) is a real system of two coupled equations for the
two-component wave functions $\phi_{1}$ and $\phi_{2}$. Thus, one
can obtain real-valued solutions for this equation, but complex-valued
solutions can also be obtained (although these do not describe a Majorana
particle) \cite{RefR}. The Majorana condition (Eq. (16)) imposed
upon $\Psi$ leads us to the following relation:
\begin{equation}
\Psi=\Psi^{*}\,(\,\Leftrightarrow\,\phi_{1}=\phi_{1}^{*}\equiv(\phi_{1})_{C}\:\,\mathrm{and}\:\,\phi_{2}=\phi_{2}^{*}\equiv(\phi_{2})_{C}\,).
\end{equation}
That is, the Majorana condition imposed on the Dirac wave function
in the Majorana representation is what implies that this wave function
must be real.

In passing, we note that the matrices originally chosen by Majorana
in his 1937 article were the following \cite{RefA}: $\hat{\alpha}_{1}'=\hat{\sigma}_{x}\otimes\hat{\sigma}_{x}$,
$\hat{\alpha}_{2}'=\hat{\sigma}_{z}\otimes\hat{1}_{2}$, $\hat{\alpha}_{3}'=\hat{\sigma}_{x}\otimes\hat{\sigma}_{z}$
and $\mathbf{\hat{\beta}'=}-\hat{\sigma}_{x}\otimes\hat{\sigma}_{y}$.
Of these four matrices, only $\hat{\alpha}_{2}'$ coincides with our
matrix $\hat{\alpha}_{2}$. The other three matrices differ from ours
by a minus sign. Incidentally, these two representations are related
by $\hat{\boldsymbol{\alpha}}'=\hat{S}\,\hat{\boldsymbol{\alpha}}\,\hat{S}^{\dagger}$,
$\hat{\beta}'=\hat{S}\,\hat{\beta}\,\hat{S}^{\dagger}$, and $\left[\,\phi_{1}'\;\phi_{2}'\,\right]^{\mathrm{T}}=\hat{S}\left[\,\phi_{1}\;\phi_{2}\,\right]^{\mathrm{T}}$,
where $\hat{S}=\hat{\sigma}_{z}\otimes\hat{1}_{2}=\hat{S}^{\dagger}$. 

\noindent \textcompwordmark{}

\noindent \textbf{In (1+1) dimensions.} We write the two-component
Dirac wave function $\Psi$ in the form given in Eq. (41), but in
this case, $\phi_{1}$ and $\phi_{2}$ are simply functions of a single
component. The Dirac equation (Eq. (3)) has the form
\begin{equation}
\mathrm{i}\hbar\frac{\partial}{\partial t}\left[\begin{array}{c}
\phi_{1}\\
\phi_{2}
\end{array}\right]=\hat{\mathrm{H}}\left[\begin{array}{c}
\phi_{1}\\
\phi_{2}
\end{array}\right]=\left[\begin{array}{cc}
0 & -\mathrm{i}\hbar c\frac{\partial}{\partial x}-\mathrm{i}(V_{\mathrm{S}}+\mathrm{m}c^{2})\\
-\mathrm{i}\hbar c\frac{\partial}{\partial x}+\mathrm{i}(V_{\mathrm{S}}+\mathrm{m}c^{2}) & 0
\end{array}\right]\left[\begin{array}{c}
\phi_{1}\\
\phi_{2}
\end{array}\right].
\end{equation}
Again, the Dirac equation in this representation is a real system
of two coupled equations for the wave functions $\phi_{1}$ and $\phi_{2}$.
However, it is precisely the Majorana condition (Eq. (16)) imposed
upon $\Psi$ that leads us to the real condition of the wave function:
\begin{equation}
\Psi=\Psi^{*}\,(\,\Leftrightarrow\,\phi_{1}=\phi_{1}^{*}\equiv(\phi_{1})_{C}\:\,\mathrm{and}\:\,\phi_{2}=\phi_{2}^{*}\equiv(\phi_{2})_{C}\,).
\end{equation}
Recently, distinct real-valued general solutions of the time-dependent
Dirac equation in Eq. (44) (i.e., subject to the constraint in Eq.
(45)), for distinct scalar potentials and borders, were constructed
\cite{RefR}. Certainly, all these solutions describe a one-dimensional
Majorana particle in its respective physical situation. 

The most general set of boundary conditions for the one-dimensional
Majorana particle inside a box in the Majorana representation was
written in detail in Ref. \cite{RefF}. This set consists of two real
one-parameter families of boundary conditions (see Eqs. (67) and (68)
of that reference). The Majorana condition in the Majorana representation
leads very easily to the Majorana condition in any other representation.
In fact, we know that wave functions in the Dirac and Majorana representations
are linked through the relation $\left[\,\phi_{1}\;\phi_{2}\,\right]^{\mathrm{T}}=\hat{S}\left[\,\varphi\;\chi\,\right]^{\mathrm{T}}$,
where the matrix $\hat{S}$ is given in Eq. (13); in addition, wave
functions in the Weyl and Majorana representations are linked through
the relation $\left[\,\phi_{1}\;\phi_{2}\,\right]^{\mathrm{T}}=\hat{S}\left[\,\varphi_{1}\;\varphi_{2}\,\right]^{\mathrm{T}}$,
where the matrix $\hat{S}$ is given in Eq. (14). Thus, by imposing
the Majorana condition (Eq. (45)) on the latter two relations, we
obtain the Majorana condition in the Dirac and Weyl representations,
i.e., Eqs. (23) and (34), respectively. Certainly, the latter general
discussion is also valid in (3+1) dimensions. 

\section{Equations for the Majorana single-particle II }

\noindent \textbf{In (3+1) dimensions.} Let us define, as Case did
\cite{RefH}, the following wave functions and matrices:
\begin{equation}
\Psi_{\pm}\equiv\frac{1}{2}\left(\hat{1}_{4}\pm\hat{\gamma}^{5}\right)\Psi
\end{equation}
and
\begin{equation}
\hat{\gamma}_{\pm}^{\mu}\equiv\frac{1}{2}\left(\hat{1}_{4}\pm\hat{\gamma}^{5}\right)\hat{\gamma}^{\mu},
\end{equation}
where the matrix $\hat{\gamma}^{5}\equiv\mathrm{i}\hat{\gamma}^{0}\hat{\gamma}^{1}\hat{\gamma}^{2}\hat{\gamma}^{3}=-\mathrm{i}\hat{\alpha}_{1}\hat{\alpha}_{2}\hat{\alpha}_{3}$
is Hermitian and satisfies the relations $(\hat{\gamma}^{5})^{2}=\hat{1}_{4}$,
and $\{\hat{\gamma}^{5},\hat{\gamma}^{\mu}\}=\hat{0}_{4}$. In addition,
$\hat{\gamma}^{5}$ satisfies the relation $\hat{S}_{C}\,(-\hat{\gamma}^{5})^{*}(\hat{S}_{C})^{-1}=\hat{\gamma}^{5}$
(i.e., $\hat{\gamma}^{5}$, just as $\hat{\gamma}^{\mu}$, satisfies
Eq. (5)), and 
\begin{equation}
\left[\frac{1}{2}\left(\hat{1}_{4}\pm\hat{\gamma}^{5}\right)\right]^{2}=\frac{1}{2}\left(\hat{1}_{4}\pm\hat{\gamma}^{5}\right)\,,\quad\mathrm{and}\quad\frac{1}{2}\left(\hat{1}_{4}\pm\hat{\gamma}^{5}\right)\frac{1}{2}\left(\hat{1}_{4}\mp\hat{\gamma}^{5}\right)=\hat{0}_{4}.
\end{equation}
Note that the charge conjugate of the wave functions in (46) verify
$(\Psi_{\pm})_{C}=(\Psi_{C})_{\mp}$. The matrix $\hat{\gamma}^{5}$
is called the chirality matrix and its eigenstates are precisely $\Psi_{+}$
(the right-chiral state), with eigenvalue $+1$, and $\Psi_{-}$ (the
left-chiral state), with eigenvalue $-1$ \cite{RefC} (the latter
two results can easily be demonstrated by multiplying Eq. (46) by
$\hat{\gamma}^{5}$ from the left). However, also note that $(\Psi_{+})_{C}$
is the eigenstate of $\hat{\gamma}^{5}$ with eigenvalue $-1$ (i.e.,
it is a left-chiral state), and $(\Psi_{-})_{C}$ is the eigenstate
of $\hat{\gamma}^{5}$ with eigenvalue $+1$ (i.e., it is a right-chiral
state). The matrices $\hat{\gamma}^{5}$ and the wave functions $\Psi_{\pm}$
in the three representations that we use in this article are shown
in Tables 1 and 3, respectively. 

First, by multiplying Eq. (1) by $\tfrac{1}{2}(\hat{1}_{4}+\hat{\gamma}^{5})$
from the left, we obtain the equation
\begin{equation}
\mathrm{i}\hat{\gamma}_{+}^{\mu}\partial_{\mu}\Psi_{-}-\frac{1}{\hbar c}(V_{\mathrm{S}}+\mathrm{m}c^{2})\hat{1}_{4}\Psi_{+}=0,
\end{equation}
and similarly, by multiplying Eq. (1) by $\tfrac{1}{2}(\hat{1}_{4}-\hat{\gamma}^{5})$,
we obtain the equation
\begin{equation}
\mathrm{i}\hat{\gamma}_{-}^{\mu}\partial_{\mu}\Psi_{+}-\frac{1}{\hbar c}(V_{\mathrm{S}}+\mathrm{m}c^{2})\hat{1}_{4}\Psi_{-}=0.
\end{equation}
Because $\Psi=\Psi_{+}+\Psi_{-}$ and $\hat{\gamma}^{\mu}=\hat{\gamma}_{+}^{\mu}+\hat{\gamma}_{-}^{\mu}$,
we have that Eqs. (49) and (50) are completely equivalent to the Dirac
equation (1). Likewise, because Eq. (1) is also satisfied by the charge-conjugate
wave function, we also have two equations that are equivalent to the
Dirac equation for $\Psi_{C}$. In effect, multiplying the latter
by $\tfrac{1}{2}(\hat{1}_{4}+\hat{\gamma}^{5})$ and $\tfrac{1}{2}(\hat{1}_{4}-\hat{\gamma}^{5})$,
we obtain 
\begin{equation}
\mathrm{i}\hat{\gamma}_{+}^{\mu}\partial_{\mu}(\Psi_{+})_{C}-\frac{1}{\hbar c}(V_{\mathrm{S}}+\mathrm{m}c^{2})\hat{1}_{4}(\Psi_{-})_{C}=0\,,\quad\mathrm{and}\quad\mathrm{i}\hat{\gamma}_{-}^{\mu}\partial_{\mu}(\Psi_{-})_{C}-\frac{1}{\hbar c}(V_{\mathrm{S}}+\mathrm{m}c^{2})\hat{1}_{4}(\Psi_{+})_{C}=0,
\end{equation}
respectively (remember that $(\Psi_{\pm})_{C}=\hat{S}_{C}\Psi_{\pm}^{*}$).
Note that because $(\Psi_{+})_{C}=(\Psi_{C})_{-}$ and $(\Psi_{-})_{C}=(\Psi_{C})_{+}$,
the wave functions $\Psi_{+}$ and $\Psi_{-}$ as well as $(\Psi_{C})_{+}$
and $(\Psi_{C})_{-}$, satisfy the same system of coupled equations,
namely, Eqs. (49) and (50) (or the system in Eq. (51)), as expected. 

In the case where $\mathrm{m}c^{2}=V_{\mathrm{S}}=0$, Eqs. (49) and
(50) are decoupled, and we have $\mathrm{i}\hat{\gamma}_{+}^{\mu}\partial_{\mu}\Psi_{-}=0$
($\Rightarrow\mathrm{i}\hat{\gamma}^{\mu}\partial_{\mu}\Psi_{-}=0$)
and $\mathrm{i}\hat{\gamma}_{-}^{\mu}\partial_{\mu}\Psi_{+}=0$ ($\Rightarrow\mathrm{i}\hat{\gamma}^{\mu}\partial_{\mu}\Psi_{+}=0$).
In the Weyl representation, the latter two four-component equations
give us the usual Weyl equations (Eq. (30)). In the same way, if we
make $\mathrm{m}c^{2}=V_{\mathrm{S}}=0$ in the system in Eq. (51),
then we obtain $\mathrm{i}\hat{\gamma}_{+}^{\mu}\partial_{\mu}(\Psi_{+})_{C}=0$
($\Rightarrow\mathrm{i}\hat{\gamma}^{\mu}\partial_{\mu}(\Psi_{+})_{C}=0$)
and $\mathrm{i}\hat{\gamma}_{-}^{\mu}\partial_{\mu}(\Psi_{-})_{C}=0$
($\Rightarrow\mathrm{i}\hat{\gamma}^{\mu}\partial_{\mu}(\Psi_{-})_{C}=0$).
Certainly, in the Weyl representation, the latter two equations also
give us the usual Weyl equations (Eq. (30)).

The Majorana condition in Eq. (16) takes the form
\begin{equation}
\Psi_{-}=(\Psi_{+})_{C}\,\,(\,\Leftrightarrow\,\Psi_{+}=(\Psi_{-})_{C}\,)
\end{equation}
(remember that $(\hat{S}_{C})^{-1}=(\hat{S}_{C})^{*}$), i.e., $\Psi_{-}=(\Psi_{C})_{-}$
($\Leftrightarrow\,\Psi_{+}=(\Psi_{C})_{+}$). Substituting the latter
$\Psi_{-}$ into Eq. (50), we obtain an equation for the four-component
wave function $\Psi_{+}$, namely,
\begin{equation}
\mathrm{i}\hat{\Gamma}^{\mu}\partial_{\mu}\Psi_{+}-\frac{1}{\hbar c}(V_{\mathrm{S}}+\mathrm{m}c^{2})\hat{1}_{4}\Psi_{+}^{*}=0,
\end{equation}
where
\begin{equation}
\hat{\Gamma}^{\mu}\equiv(\hat{S}_{C})^{*}\,\hat{\gamma}_{-}^{\mu}\,,\quad\mathrm{with}\quad(\hat{\Gamma}^{\mu})^{*}\,\hat{\Gamma}^{\nu}+(\hat{\Gamma}^{\nu})^{*}\,\hat{\Gamma}^{\mu}=-2g^{\mu\nu}\frac{1}{2}\left(\hat{1}_{4}+\hat{\gamma}^{5}\right)
\end{equation}
(the equation for $\Psi_{+}$ that results after making the latter
substitution but into Eq. (49) is absolutely equivalent to Eq. (53)). 

Alternatively, substituting the wave function $\Psi_{+}$ from Eq.
(52) into Eq. (49), we obtain an equation for the four-component wave
function $\Psi_{-}$, namely,
\begin{equation}
\mathrm{i}\hat{\Lambda}^{\mu}\partial_{\mu}\Psi_{-}-\frac{1}{\hbar c}(V_{\mathrm{S}}+\mathrm{m}c^{2})\hat{1}_{4}\Psi_{-}^{*}=0,
\end{equation}
where
\begin{equation}
\hat{\Lambda}^{\mu}\equiv(\hat{S}_{C})^{*}\,\hat{\gamma}_{+}^{\mu}\,,\quad\mathrm{with}\quad(\hat{\Lambda}^{\mu})^{*}\,\hat{\Lambda}^{\nu}+(\hat{\Lambda}^{\nu})^{*}\,\hat{\Lambda}^{\mu}=-2g^{\mu\nu}\frac{1}{2}\left(\hat{1}_{4}-\hat{\gamma}^{5}\right)
\end{equation}
(again, the equation for $\Psi_{-}$ that results after making the
latter substitution but into Eq. (50) is absolutely equivalent to
Eq. (55)). Naturally, by imposing the Majorana condition (Eq. (52))
upon the equations in (51), we again obtain Eqs. (53) and (55). 

On the other hand, making $\mathrm{m}c^{2}=V_{\mathrm{S}}=0$ in Eq.
(53) leads us to the relation $\mathrm{i}\hat{\gamma}_{-}^{\mu}\partial_{\mu}\Psi_{+}=0$
($\Rightarrow\mathrm{i}\hat{\gamma}^{\mu}\partial_{\mu}\Psi_{+}=0$),
and as can be seen in Eq. (51), we also have $\mathrm{i}\hat{\gamma}_{-}^{\mu}\partial_{\mu}(\Psi_{-})_{C}=0$
($\Rightarrow\mathrm{i}\hat{\gamma}^{\mu}\partial_{\mu}(\Psi_{-})_{C}=0$),
but also in this case, we have $\Psi_{+}=(\Psi_{-})_{C}$ (this is
due to the Majorana condition). Similarly, making $\mathrm{m}c^{2}=V_{\mathrm{S}}=0$
in Eq. (55) leads us to the relation $\mathrm{i}\hat{\gamma}_{+}^{\mu}\partial_{\mu}\Psi_{-}=0$
($\Rightarrow\mathrm{i}\hat{\gamma}^{\mu}\partial_{\mu}\Psi_{-}=0$),
but from Eq. (51) we also have $\mathrm{i}\hat{\gamma}_{+}^{\mu}\partial_{\mu}(\Psi_{+})_{C}=0$
($\Rightarrow\mathrm{i}\hat{\gamma}^{\mu}\partial_{\mu}(\Psi_{+})_{C}=0$),
where $\Psi_{-}=(\Psi_{+})_{C}$ (this is also due to the Majorana
condition). 

To obtain the four-component wave function that describes the Majorana
particle, namely, $\Psi=\Psi_{+}+\Psi_{-}$, it is sufficient to solve
the equation for $\Psi_{+}$ (Eq. (53)), and then, from this solution,
and using the Majorana condition in (52), one obtains $\Psi_{-}$.
Alternatively, one could also solve the equation for $\Psi_{-}$ (Eq.
(55)), and then, from this solution, and using the Majorana condition
in (52), one obtains $\Psi_{+}$. Note that, in the former case, $\Psi=\Psi_{+}+(\Psi_{+})_{C}$,
and therefore, $\Psi=\Psi_{C}$ (remember that $((\Psi_{+})_{C})_{C}=\Psi_{+}$);
similarly, in the latter case, $\Psi=(\Psi_{-})_{C}+\Psi_{-}$, and
therefore, $\Psi=\Psi_{C}$ (remember that $((\Psi_{-})_{C})_{C}=\Psi_{-}$),
as expected. Clearly, the four-component wave function $\Psi$ depends
only on the solution of Eq. (53) (or of Eq. (55)); thus, we can consider
that Eq. (53) (or Eq. (55)) alone models the Majorana particle in
(3+1) dimensions and in a form independent of the choice of representation.

Certainly, the above-mentioned procedure to obtain $\Psi$ is general,
but in each representation, it has its own particularity. In relation
to this, we can now obtain different results. In the rest of this
subsection, we make full use of Tables 3, 4 and 5. First, in the Weyl
representation, the covariant four-component equation for $\Psi_{+}=\left[\,\varphi_{1}\;\,0\,\right]^{\mathrm{T}}$
(Eq. (53)) leads us to the following explicitly covariant two-component
equation for the two-component wave function $\varphi_{1}$:
\begin{equation}
\hat{\eta}^{\mu}\partial_{\mu}\varphi_{1}-\frac{1}{\hbar c}(V_{\mathrm{S}}+\mathrm{m}c^{2})\hat{1}_{2}\varphi_{1}^{*}=0,
\end{equation}
where the matrices $\hat{\eta}^{0}=-\mathrm{i}\hat{\sigma}_{y}$,
$\hat{\eta}^{1}=-\hat{\sigma}_{z}$, $\hat{\eta}^{2}=-\mathrm{i}\hat{1}_{2}$,
and $\hat{\eta}^{3}=\hat{\sigma}_{x}$, satisfy the relation
\begin{equation}
(\hat{\eta}^{\mu})^{*}\,\hat{\eta}^{\nu}+(\hat{\eta}^{\nu})^{*}\,\hat{\eta}^{\mu}=-2g^{\mu\nu}\hat{1}_{2}
\end{equation}
(this last relation arises from Eq. (54)). After multiplying Eq. (57)
by $-\hat{\sigma}_{y}$, this equation takes an alternative form,
namely, 
\begin{equation}
\mathrm{i}\hat{\sigma}^{\mu}\partial_{\mu}\varphi_{1}+\frac{1}{\hbar c}(V_{\mathrm{S}}+\mathrm{m}c^{2})\hat{\sigma}_{y}\varphi_{1}^{*}=0,
\end{equation}
where $\hat{\sigma}^{0}=\hat{1}_{2}$, $\hat{\sigma}^{1}=\hat{\sigma}_{x}$,
$\hat{\sigma}^{2}=\hat{\sigma}_{y}$, and $\hat{\sigma}^{3}=\hat{\sigma}_{z}$
(or, as it is commonly written, $\hat{\sigma}^{\mu}=(\hat{1}_{2},+\hat{\boldsymbol{\sigma}})$)
\cite{RefH}. Equation (59) is precisely Eq. (28), as expected. Now,
if we use the Majorana condition (Eq. (52)), we can obtain $\Psi_{-}=\left[\,0\;\,\varphi_{2}\,\right]^{\mathrm{T}}$
from $\Psi_{+}=\left[\,\varphi_{1}\;\,0\,\right]^{\mathrm{T}}$, and
the result is $\Psi_{-}=\left[\,0\;\,\hat{\sigma}_{y}\,\varphi_{1}^{*}\,\right]^{\mathrm{T}}$
(which is in agreement with the result in Eq. (27)). Finally, we can
write the four-component wave function for the Majorana particle,
namely, $\Psi=\Psi_{+}+\Psi_{-}=\left[\,\varphi_{1}\;\,\hat{\sigma}_{y}\,\varphi_{1}^{*}\,\right]^{\mathrm{T}}$.
It is clear that this four-component solution is dependent only on
the two-component complex wave function $\varphi_{1}$, which is the
solution of Eq. (59), i.e., here, we have only four independent real
quantities. Because we have $\hat{\gamma}^{5}\Psi_{+}=(+1)\Psi_{+}$,
Eq. (59) is referred to as the right-chiral two-component Majorana
equation.

Similarly, the covariant four-component equation for $\Psi_{-}=\left[\,0\;\,\varphi_{2}\,\right]^{\mathrm{T}}$
(Eq. (55)) leads us to the following explicitly covariant two-component
equation for the two-component wave function $\varphi_{2}$:
\begin{equation}
\hat{\xi}^{\mu}\partial_{\mu}\varphi_{2}-\frac{1}{\hbar c}(V_{\mathrm{S}}+\mathrm{m}c^{2})\hat{1}_{2}\varphi_{2}^{*}=0,
\end{equation}
where the matrices $\hat{\xi}^{0}=-\hat{\eta}^{0}$, $\hat{\xi}^{j}=\hat{\eta}^{j}$,
with $j=1,2,3$, also satisfy Eq. (58) (in this case, the latter relation
arises from Eq. (56)). Multiplying Eq. (60) by $\hat{\sigma}_{y}$,
this equation takes the alternative form 
\begin{equation}
\mathrm{i}\hat{\bar{\sigma}}^{\mu}\partial_{\mu}\varphi_{2}-\frac{1}{\hbar c}(V_{\mathrm{S}}+\mathrm{m}c^{2})\hat{\sigma}_{y}\varphi_{2}^{*}=0,
\end{equation}
where $\hat{\bar{\sigma}}^{0}=\hat{\sigma}^{0}$, $\hat{\bar{\sigma}}^{1}=-\hat{\sigma}^{1}$,
$\hat{\bar{\sigma}}^{2}=-\hat{\sigma}^{2}$, and $\hat{\bar{\sigma}}^{3}=-\hat{\sigma}^{3}$
(i.e., $\hat{\bar{\sigma}}^{\mu}=(\hat{1}_{2},-\hat{\boldsymbol{\sigma}})$).
Equation (61) is precisely Eq. (29), as expected. Again, if we use
the Majorana condition (Eq. (52)), we can obtain $\Psi_{+}=\left[\,\varphi_{1}\;\,0\,\right]^{\mathrm{T}}$
from $\Psi_{-}=\left[\,0\;\,\varphi_{2}\,\right]^{\mathrm{T}}$, and
the result is $\Psi_{+}=\left[\,-\hat{\sigma}_{y}\,\varphi_{2}^{*}\;\,0\,\right]^{\mathrm{T}}$
(which is in agreement with the result in Eq. (27)). Thus, we can
write the four-component wave function for the Majorana particle,
namely, $\Psi=\Psi_{+}+\Psi_{-}=\left[\,-\hat{\sigma}_{y}\,\varphi_{2}^{*}\;\,\varphi_{2}\,\right]^{\mathrm{T}}$.
The latter four-component solution depends only on the two-component
complex wave function $\varphi_{2}$, which is the solution of Eq.
(61), i.e., here, we have only four independent real quantities, as
expected for a Majorana particle. Because we have $\hat{\gamma}^{5}\Psi_{-}=(-1)\Psi_{-}$,
Eq. (61) is referred to as the left-chiral two-component Majorana
equation.

In summary, Eq. (59) alone can be considered as a Majorana equation
for the Majorana particle, even for a particular type of Majorana
particle. Likewise, Eq. (61) alone can also be considered as a Majorana
equation for the Majorana particle, even for a Majorana particle different
from the previous one (for example, with a different mass). Thus,
Eqs. (59) and (61), although similar, are non-equivalent two-component
equations. Specifically, this is because $\varphi_{1}$ and $\varphi_{2}$
transform in two precise and different ways under a Lorentz boost,
i.e., they transform according to two inequivalent representations
of the Lorentz group \cite{RefI}. Certainly, Eqs. (59) and (61) tend
toward the pair of Weyl equations when $\mathrm{m}c^{2}=V_{\mathrm{S}}=0$
(Eq. (30)). Equations (59) and (61) comprise the so-called two-component
theory of Majorana particles \cite{RefH}. 

Again, in the Weyl representation that we have considered in our paper
($\hat{\gamma}^{0}=\hat{\beta}=-\hat{\sigma}_{x}\otimes\hat{1}_{2}$,
$\hat{\boldsymbol{\gamma}}=\hat{\beta}\hat{\boldsymbol{\alpha}}=+\mathrm{i}\hat{\sigma}_{y}\otimes\hat{\boldsymbol{\sigma}}$,
and $\hat{\gamma}^{5}=+\hat{\sigma}_{z}\otimes\hat{1}_{2}$), we used
$\hat{S}_{C}=-\hat{\gamma}^{2}=-\mathrm{i}\hat{\sigma}_{y}\otimes\hat{\sigma}_{y}$,
but this is only because we decided to derive this result from Eq.
(8) (with $\hat{S}$ given by Eq. (10)). We could, for example, write
$\hat{S}_{C}=-\mathrm{i}\hat{\gamma}^{2}=+\hat{\sigma}_{y}\otimes\hat{\sigma}_{y}$.
In the latter case, the equations for $\varphi_{1}$ and $\varphi_{2}$
are simply Eqs. (59) and (61) with the following replacement: $\hat{\sigma}_{y}\rightarrow+\mathrm{i}\hat{\sigma}_{y}$,
namely, 
\begin{equation}
\mathrm{i}\hat{\sigma}^{\mu}\partial_{\mu}\varphi_{1}+\frac{1}{\hbar c}(V_{\mathrm{S}}+\mathrm{m}c^{2})\mathrm{i}\hat{\sigma}_{y}\varphi_{1}^{*}=0,
\end{equation}
and 
\begin{equation}
\mathrm{i}\hat{\bar{\sigma}}^{\mu}\partial_{\mu}\varphi_{2}-\frac{1}{\hbar c}(V_{\mathrm{S}}+\mathrm{m}c^{2})\mathrm{i}\hat{\sigma}_{y}\varphi_{2}^{*}=0.
\end{equation}

Equations (62) and (63) are essentially Eqs. (71) and (70) given in
Ref. \cite{RefI}, respectively. In effect, as already mentioned before,
in that remarkable reference the matrices $\hat{\boldsymbol{\alpha}}'=+\hat{\sigma}_{z}\otimes\hat{\boldsymbol{\sigma}}$,
$\hat{\gamma}^{0}\,'=\hat{\beta}'=+\hat{\sigma}_{x}\otimes\hat{1}_{2}$,
and $\hat{\boldsymbol{\gamma}}'=\hat{\beta}'\hat{\boldsymbol{\alpha}}'=-\mathrm{i}\hat{\sigma}_{y}\otimes\hat{\boldsymbol{\sigma}}$,
with $\left[\,\varphi_{1}'\;\varphi_{2}'\,\right]^{\mathrm{T}}\equiv\left[\,\tilde{\psi}\;\,\psi\,\right]^{\mathrm{T}}$,
were considered as the Weyl representation. These matrices and those
used by us are related through the relations $\hat{\boldsymbol{\alpha}}'=\hat{S}\,\hat{\boldsymbol{\alpha}}\,\hat{S}^{\dagger}$,
$\hat{\beta}'=\hat{S}\,\hat{\beta}\,\hat{S}^{\dagger}$, $\hat{\boldsymbol{\gamma}}'=\hat{S}\,\hat{\boldsymbol{\gamma}}\,\hat{S}^{\dagger}$,
and $\left[\,\varphi_{1}'\;\varphi_{2}'\,\right]^{\mathrm{T}}=\hat{S}\left[\,\varphi_{1}\;\varphi_{2}\,\right]^{\mathrm{T}}$,
where $\hat{S}=\hat{\sigma}_{z}\otimes\hat{1}_{2}=\hat{S}^{\dagger}$.
In addition, the charge-conjugation matrices, $\hat{S}_{C}'=-\mathrm{i}\hat{\gamma}^{2}\,'=-\hat{\sigma}_{y}\otimes\hat{\sigma}_{y}$
and $\hat{S}_{C}=-\mathrm{i}\hat{\gamma}^{2}=+\hat{\sigma}_{y}\otimes\hat{\sigma}_{y}$,
are related by means of Eq. (7). Thus, Eq. (62) for $\varphi_{1}=\varphi_{1}'\equiv\tilde{\psi}$,
with $V_{\mathrm{S}}=0$, is Eq. (71) of Ref. \cite{RefI}, and Eq.
(63) for $\varphi_{2}=-\varphi_{2}'\equiv-\psi$, also with $V_{\mathrm{S}}=0$,
is Eq. (70) of the same reference. Also, in Ref. \cite{RefI}, the
former equation was appropriately named the right-chiral two-component
Majorana equation, and the latter was named the left-chiral two-component
Majorana equation. 

Unsurprisingly, Eqs. (62) and (63) for $\varphi_{1}$ and $\varphi_{2}$
can be written jointly in the form
\begin{equation}
\mathrm{i}\hat{\gamma}^{\mu}\partial_{\mu}\Psi-\frac{1}{\hbar c}(V_{\mathrm{S}}+\mathrm{m}c^{2})\hat{1}_{4}\Psi_{C}=0,
\end{equation}
where $\Psi=\left[\,\varphi_{1}\;\varphi_{2}\,\right]^{\mathrm{T}}$
and $\Psi_{C}\equiv\hat{S}_{C}\,\Psi^{*}$ with $\hat{S}_{C}=-\mathrm{i}\hat{\gamma}^{2}=+\hat{\sigma}_{y}\otimes\hat{\sigma}_{y}$.
Specifically, Eq. (64) is the (four-component) Majorana equation with
a scalar potential (see the discussion on this equation in the introduction).
However, if this equation is considered to describe a Majorana particle
with a four-component wave function, $\Psi=\left[\,\varphi_{1}\;\varphi_{2}\,\right]^{\mathrm{T}}$,
it should be remembered that, due to the Majorana condition, $\Psi=\Psi_{C}$,
$\varphi_{1}$ and $\varphi_{2}$ are not independent two-component
wave functions. Therefore, in this case, it would be sufficient to
solve just one of the two two-component Majorana equations, and then,
with the relation between $\varphi_{1}$ and $\varphi_{2}$, we could
reconstruct the entire wave function $\Psi$. However, if Eq. (64)
is considered to describe a Majoranon \cite{RefG,RefS}, then the
two two-component Majorana equations must be solved, the solutions
of which are simply the top and bottom components of the wave function
$\Psi$ in Eq. (64). The latter result is somewhat unexpected. Similarly,
Eqs. (62) and (63) for $\tilde{\psi}$ and $\psi$ in Ref. \cite{RefI},
respectively, can also be combined into Eq. (64). In this case, we
write Eq. (64) with the following replacements: $\Psi\rightarrow\Psi'\equiv\Psi_{M}$,
$\Psi_{C}\rightarrow\Psi_{C}'\equiv\Psi_{M}^{c}$, and $\hat{\gamma}^{\mu}\rightarrow\hat{\gamma}^{\mu}\,'\equiv\hat{\gamma}^{\mu}$,
where $\Psi'=\left[\,\tilde{\psi}\;\,\psi\,\right]^{\mathrm{T}}$
and $\Psi_{C}'\equiv\hat{S}_{C}'\,\Psi'^{*}$ with $\hat{S}_{C}'=-\mathrm{i}\hat{\gamma}^{2}\,'=-\hat{\sigma}_{y}\otimes\hat{\sigma}_{y}$.
In the present case, Eq. (64), with $V_{\mathrm{S}}=0$, is Eq. (123)
of Ref. \cite{RefI}. 

The same Weyl representation used in Ref. \cite{RefI} was used in
Ref. \cite{RefT}, but here, $\hat{S}_{C}'=+\mathrm{i}\hat{\gamma}^{2}\,'=+\hat{\sigma}_{y}\otimes\hat{\sigma}_{y}$
was chosen. Thus, in this case, the Majorana equations for $\varphi_{1}'$
and $\varphi_{2}'\,(=+\mathrm{i}\hat{\sigma}_{y}\,\varphi_{1}^{*}\,')$
can be obtained from Eqs. (62) and (63) by making the following substitutions:
$\varphi_{1}\rightarrow\varphi_{1}'$, $\varphi_{2}\rightarrow\varphi_{2}'$,
and $\hat{\sigma}_{y}\rightarrow-\hat{\sigma}_{y}$. The equation
for $\varphi_{1}'\equiv\phi$, with $+\mathrm{i}\hat{\sigma}_{y}\,\phi^{*}\equiv\hat{\mathrm{S}}\phi$,
and $V_{\mathrm{S}}=0$, is Eq. (13) of Ref. \cite{RefT}, and the
equation for $\varphi_{2}'=+\mathrm{i}\hat{\sigma}_{y}\,\phi^{*}\equiv\hat{\mathrm{S}}\phi$,
with $V_{\mathrm{S}}=0$, is Eq. (14) of the same reference. Incidentally,
by linearizing the standard relativistic energy\textendash{}momentum
relation, and without recourse to the Dirac equation, a good derivation
of the two-component Majorana equation for $\varphi_{1}'\equiv\phi$,
with $V_{\mathrm{S}}=0$, was obtained in Ref. \cite{RefT}. 

The following matrices are also a very common choice when introducing
the Weyl representation: $\hat{\gamma}^{0}\,'=\hat{\beta}'=+\hat{\sigma}_{x}\otimes\hat{1}_{2}$,
$\hat{\boldsymbol{\gamma}}'=\hat{\beta}'\hat{\boldsymbol{\alpha}}'=+\mathrm{i}\hat{\sigma}_{y}\otimes\hat{\boldsymbol{\sigma}}$,
and $\hat{\gamma}^{5}\,'=-\hat{\sigma}_{z}\otimes\hat{1}_{2}$, with
the respective wave function written in the form $\left[\,\varphi_{1}'\;\varphi_{2}'\,\right]^{\mathrm{T}}$.
These matrices and those used by us are related as follows: $\hat{\gamma}^{\mu}\,'=\hat{S}\,\hat{\gamma}^{\mu}\hat{S}^{\dagger}$,
etc, and $\left[\,\varphi_{1}'\;\varphi_{2}'\,\right]^{\mathrm{T}}=\hat{S}\left[\,\varphi_{1}\;\varphi_{2}\,\right]^{\mathrm{T}}$,
where $\hat{S}=\hat{\sigma}_{y}\otimes\hat{1}_{2}=\hat{S}^{\dagger}$.
In addition, by substituting the latter matrix, and $\hat{S}_{C}=-\mathrm{i}\hat{\gamma}^{2}=+\hat{\sigma}_{y}\otimes\hat{\sigma}_{y}$
into Eq. (7), we obtain $\hat{S}_{C}'=-\hat{\sigma}_{y}\otimes\hat{\sigma}_{y}=+\mathrm{i}\hat{\gamma}^{2}\,'$
(which is a typical choice when considering this Weyl representation).
Then, the equations for $\varphi_{1}'$ and $\varphi_{2}'$ are given
by 
\begin{equation}
\mathrm{i}\hat{\bar{\sigma}}^{\mu}\partial_{\mu}\varphi_{1}'+\frac{1}{\hbar c}(V_{\mathrm{S}}+\mathrm{m}c^{2})\mathrm{i}\hat{\sigma}_{y}\varphi_{1}'^{*}=0,
\end{equation}
and 
\begin{equation}
\mathrm{i}\hat{\sigma}^{\mu}\partial_{\mu}\varphi_{2}'-\frac{1}{\hbar c}(V_{\mathrm{S}}+\mathrm{m}c^{2})\mathrm{i}\hat{\sigma}_{y}\varphi_{2}'^{*}=0.
\end{equation}
Equation (65) for $\varphi_{1}'\equiv\omega$, with $V_{\mathrm{S}}=0$,
is precisely Eq. (107) of Ref. \cite{RefC}, but this equation is
the left-chiral two-component Majorana equation because, in this Weyl
representation, one has $\hat{\gamma}^{5}\,'=-\hat{\sigma}_{z}\otimes\hat{1}_{2}$.
We mention in passing that the reference by Pal is a great tutorial
article that addresses in detail the key connections between the Dirac,
Majorana, and Weyl fields in (3+1) dimensions. Likewise, Eq. (65)
for $\varphi_{1}'\equiv\nu$, with $V_{\mathrm{S}}=0$, is just Eq.
(4.93) of the renowned book by Mohapatra and Pal \cite{RefU}. Note
that in the latter reference, the following notation was used: $\hat{\sigma}_{\mu}=(\hat{1}_{2},+\hat{\boldsymbol{\sigma}})\Rightarrow\hat{\sigma}^{\mu}=(\hat{1}_{2},-\hat{\boldsymbol{\sigma}})$,
and $\hat{\bar{\sigma}}_{\mu}=(\hat{1}_{2},-\hat{\boldsymbol{\sigma}})\Rightarrow\hat{\bar{\sigma}}^{\mu}=(\hat{1}_{2},+\hat{\boldsymbol{\sigma}})$,
instead of the most common notation that we use in this article (see
Eqs. (59) and (61)). Moreover, in this reference, the Majorana condition
was written as $\Psi=\exp(-\mathrm{i}\delta)\Psi_{C}$ instead of
as $\Psi=\Psi_{C}$, which is our choice. Likewise, Eq. (65) for $\varphi_{1}'\equiv\chi$,
with $V_{\mathrm{S}}=0$, is Eq. (9) of Ref. \cite{RefV}. In the
latter great reference, a coupled system of two left-chiral Majorana
equations was constructed and used to study neutrino oscillations
for two Majorana neutrino flavor states. 

Second, in the Dirac representation, the covariant four-component
equation for $\Psi_{+}$ (Eq. (53)) leads us to the covariant two-component
Eq. (57) with the following replacement: $\varphi_{1}\rightarrow\varphi+\chi$.
Likewise, the Majorana condition in Eq. (52) leads us to Eq. (27)
with the latter replacement plus the following: $\varphi_{2}\rightarrow-\varphi+\chi$,
namely, $-\varphi+\chi=\hat{\sigma}_{y}(\varphi+\chi)^{*}$ (Eq. (19)).
Remember that the four-component wave functions in the Dirac and Weyl
representations are related through the relation $\left[\,\varphi_{1}\;\varphi_{2}\,\right]^{\mathrm{T}}=\hat{S}\left[\,\varphi\;\chi\,\right]^{\mathrm{T}}$,
where the matrix $\hat{S}$ is given in Eq. (11). Thus, from Eq. (53),
one obtains the two-component wave function $\varphi+\chi$, from
which one can construct $\Psi_{+}$, and using the Majorana condition,
one obtains $-\varphi+\chi$ , from which one can construct $\Psi_{-}$
(see Table 3). Finally, the four-component wave function for the Majorana
particle, namely, $\Psi=\Psi_{+}+\Psi_{-}=\left[\,\varphi\;\,\chi\,\right]^{\mathrm{T}}$,
can be written immediately. Similarly, the covariant four-component
equation for $\Psi_{-}$ (Eq. (55)) leads us to Eq. (60) with the
following replacement: $\varphi_{2}\rightarrow-\varphi+\chi$. Thus,
from Eq. (55), one obtains the two-component wave function $-\varphi+\chi$,
from which one can construct $\Psi_{-}$, and using the Majorana condition
one obtains $\varphi+\chi$ , from which one can construct $\Psi_{+}$
(see Table 3). Finally, the four-component wave function for the Majorana
particle can be written immediately. 

Alternatively, adding and subtracting the former equations that result
from Eqs. (53) and (55) and once again using (conveniently) the Majorana
condition given in Eq. (19), one obtains an equation for the two-component
wave function $\varphi$, namely, 
\begin{equation}
\hat{\eta}^{0}\partial_{0}\varphi+\sum_{k=1}^{3}\hat{\eta}^{k}\partial_{k}\left(\hat{\sigma}_{y}\,\varphi^{*}\right)+\frac{1}{\hbar c}(V_{\mathrm{S}}+\mathrm{m}c^{2})\hat{\sigma}_{y}\,\varphi=0,
\end{equation}
and another equation for the two-component wave function $\chi$,
namely,
\begin{equation}
\hat{\xi}^{0}\partial_{0}\chi+\sum_{k=1}^{3}\hat{\xi}^{k}\partial_{k}\left(\hat{\sigma}_{y}\,\chi^{*}\right)+\frac{1}{\hbar c}(V_{\mathrm{S}}+\mathrm{m}c^{2})\hat{\sigma}_{y}\,\chi=0.
\end{equation}
Certainly, Eq. (67) leads to Eq. (20), and Eq. (68) leads to Eq. (21).
Likewise, from the solution of Eq. (67) or Eq. (68), and properly
using in each case the Majorana condition (Eq. (19)), one can obtain
the respective four-component wave function $\Psi=\left[\,\varphi\;\,\chi\,\right]^{\mathrm{T}}$. 

Third, in the Majorana representation, the covariant four-component
equation for $\Psi_{+}$ (Eq. (53)) is precisely Eq. (50), and the
covariant four-component equation for $\Psi_{-}$ (Eq. (55)) is precisely
Eq. (49); additionally, the latter equation is the complex conjugate
of the former equation. This is shown by the following results. Remember
that, in this representation, $\hat{S}_{C}=\hat{1}_{4}$; therefore,
$\hat{\Gamma}^{\mu}=\hat{\gamma}_{-}^{\mu}$, $\hat{\Lambda}^{\mu}=\hat{\gamma}_{+}^{\mu}$,
and, from the Majorana condition in Eq. (52), we have $\Psi_{-}=\Psi_{+}^{*}$
(and therefore, $\Psi=\Psi_{+}+\Psi_{-}=\Psi^{*}$, as expected).
In this representation, one also has $\hat{\gamma}^{\mu}=-(\hat{\gamma}^{\mu})^{*}$
and $\hat{\gamma}^{5}=-(\hat{\gamma}^{5})^{*}$, and therefore, $\hat{\gamma}_{-}^{\mu}=-(\hat{\gamma}_{+}^{\mu})^{*}$.
Thus, in the Majorana representation, the equation for the Majorana
particle is essentially Eq. (50), where $\hat{\gamma}_{-}^{\mu}=-(\hat{\gamma}_{+}^{\mu})^{*}$
and $\Psi_{-}=\Psi_{+}^{*}$ (in fact, substituting the latter relations
in the complex conjugate equation of Eq. (50), one obtains Eq. (49)).
Specifically, Eq. (50) leads us to Eq. (57) with the following replacement:
$\varphi_{1}\rightarrow(\hat{1}_{2}+\hat{\sigma}_{y})\phi_{1}-(\hat{1}_{2}-\hat{\sigma}_{y})\phi_{2}$.
Remember that the four-component wave functions in the Majorana and
Weyl representations are related by $\left[\,\varphi_{1}\;\varphi_{2}\,\right]^{\mathrm{T}}=\hat{S}^{-1}\left[\,\phi_{1}\;\phi_{2}\,\right]^{\mathrm{T}}$,
where the matrix $\hat{S}$ is given in Eq. (10) (additionally, $\varphi_{1}$
and $\varphi_{2}$ are related by Eq. (27), i.e., the Majorana condition,
which implies that $\phi_{1}$ and $\phi_{2}$ are real-valued wave
functions, as expected). Finally, the equation obtained here and its
complex conjugate can be written in the form given in Eq. (42). 

\noindent \textcompwordmark{}

\noindent \textbf{In (1+1) dimensions.} Let us introduce the following
wave functions and matrices: 
\begin{equation}
\Psi_{\pm}\equiv\frac{1}{2}\left(\hat{1}_{2}\pm\hat{\Gamma}^{5}\right)\Psi\,,\quad\mathrm{and}\quad\hat{\gamma}_{\pm}^{\mu}\equiv\frac{1}{2}\left(\hat{1}_{2}\pm\hat{\Gamma}^{5}\right)\hat{\gamma}^{\mu},
\end{equation}
where the matrix $\hat{\Gamma}^{5}\equiv-\mathrm{i}\hat{\gamma}^{5}$
is Hermitian because $\hat{\gamma}^{5}\equiv\mathrm{i}\hat{\gamma}^{0}\hat{\gamma}^{1}=\mathrm{i}\hat{\alpha}$
is anti-Hermitian, and satisfies the relations $(\hat{\Gamma}^{5})^{2}=\hat{1}_{2}$
and $\{\hat{\Gamma}^{5},\hat{\gamma}^{\mu}\}=\hat{0}_{2}$. In addition,
$\hat{\Gamma}^{5}$ satisfies the relation $\hat{S}_{C}\,(\hat{\Gamma}^{5})^{*}(\hat{S}_{C})^{-1}=\hat{\Gamma}^{5}$
(which is different from the analogous relation that satisfies $\hat{\gamma}^{5}$
in (3+1) dimensions), and
\begin{equation}
\left[\frac{1}{2}\left(\hat{1}_{2}\pm\hat{\Gamma}^{5}\right)\right]^{2}=\frac{1}{2}\left(\hat{1}_{2}\pm\hat{\Gamma}^{5}\right)\,,\quad\mathrm{and}\quad\frac{1}{2}\left(\hat{1}_{2}\pm\hat{\Gamma}^{5}\right)\frac{1}{2}\left(\hat{1}_{2}\mp\hat{\Gamma}^{5}\right)=\hat{0}_{2}.
\end{equation}
Simply note that in (1+1) dimensions, $\hat{\Gamma}^{5}=\hat{\alpha}$
acts similar to the standard fifth gamma matrix in (3+1) dimensions,
i.e., as the chirality matrix \cite{RefW,RefX}. However, in this
case, the charge conjugate of the wave functions in (69) verify $(\Psi_{\pm})_{C}=(\Psi_{C})_{\pm}$.
Thus, although it is verified that $\hat{\Gamma}^{5}\Psi_{\pm}=(\pm1)\Psi_{\pm}$,
we now have that $\hat{\Gamma}^{5}(\Psi_{\pm})_{C}=(\pm1)(\Psi_{\pm})_{C}$,
i.e., $\Psi_{\pm}$ and $(\Psi_{\pm})_{C}$ are eigenstates of $\hat{\Gamma}^{5}$
with eigenvalues $\pm1$. The matrices $\hat{\Gamma}^{5}$ and the
wave functions $\Psi_{\pm}$ in each of the three representations
that we use in this article are shown in Tables 2 and 6, respectively. 

First, note that by multiplying the Dirac equation in Eq. (1) (but
particularized to the case of (1+1) dimensions) by $\tfrac{1}{2}(\hat{1}_{2}+\hat{\Gamma}^{5})$
from the left, we obtain the equation
\begin{equation}
\mathrm{i}\hat{\gamma}_{+}^{\mu}\partial_{\mu}\Psi_{-}-\frac{1}{\hbar c}(V_{\mathrm{S}}+\mathrm{m}c^{2})\hat{1}_{2}\Psi_{+}=0,
\end{equation}
and similarly, multiplying Eq. (1) by $\tfrac{1}{2}(\hat{1}_{2}-\hat{\Gamma}^{5})$,
we obtain the equation
\begin{equation}
\mathrm{i}\hat{\gamma}_{-}^{\mu}\partial_{\mu}\Psi_{+}-\frac{1}{\hbar c}(V_{\mathrm{S}}+\mathrm{m}c^{2})\hat{1}_{2}\Psi_{-}=0.
\end{equation}
The latter pair of equations is completely equivalent to the Dirac
equation and similar to the pair of Eqs. (49) and (50) in (3+1) dimensions.
However, only in the present case, the gamma matrices in Eqs. (69)
and (70) satisfy the relations 
\begin{equation}
\hat{\gamma}_{\pm}^{\mu}\,\hat{\gamma}_{\mp}^{\nu}+\hat{\gamma}_{\pm}^{\nu}\,\hat{\gamma}_{\mp}^{\mu}=2g^{\mu\nu}\frac{1}{2}\left(\hat{1}_{2}\pm\hat{\Gamma}^{5}\right),
\end{equation}
and $\{\hat{\gamma}_{+}^{\mu},\hat{\gamma}_{+}^{\nu}\}=\{\hat{\gamma}_{-}^{\mu},\hat{\gamma}_{-}^{\nu}\}=\hat{0}_{2}$.
The charge-conjugate wave function also satisfies the Dirac equation;
thus, we also have two equations equivalent to the latter equation.
Specifically, by multiplying the Dirac equation for $\Psi_{C}$ by
$\tfrac{1}{2}(\hat{1}_{2}+\hat{\Gamma}^{5})$ and $\tfrac{1}{2}(\hat{1}_{2}-\hat{\Gamma}^{5})$,
we obtain
\begin{equation}
\mathrm{i}\hat{\gamma}_{+}^{\mu}\partial_{\mu}(\Psi_{-})_{C}-\frac{1}{\hbar c}(V_{\mathrm{S}}+\mathrm{m}c^{2})\hat{1}_{2}(\Psi_{+})_{C}=0\,,\quad\mathrm{and}\quad\mathrm{i}\hat{\gamma}_{-}^{\mu}\partial_{\mu}(\Psi_{+})_{C}-\frac{1}{\hbar c}(V_{\mathrm{S}}+\mathrm{m}c^{2})\hat{1}_{2}(\Psi_{-})_{C}=0,
\end{equation}
respectively (remember that $(\Psi_{\pm})_{C}=\hat{S}_{C}\Psi_{\pm}^{*}$).
Note that just as $\Psi_{-}$ and $\Psi_{+}$ satisfy Eqs. (71) and
(72), $(\Psi_{C})_{-}$ and $(\Psi_{C})_{+}$ also satisfy them (this
is because $(\Psi_{\pm})_{C}=(\Psi_{C})_{\pm}$). In the case of $\mathrm{m}c^{2}=V_{\mathrm{S}}=0$,
we obtain $\mathrm{i}\hat{\gamma}_{+}^{\mu}\partial_{\mu}\Psi_{-}=\mathrm{i}\hat{\gamma}_{+}^{\mu}\partial_{\mu}(\Psi_{-})_{C}=0$
($\Rightarrow\mathrm{i}\hat{\gamma}^{\mu}\partial_{\mu}\Psi_{-}=\mathrm{i}\hat{\gamma}^{\mu}\partial_{\mu}(\Psi_{-})_{C}=0$)
and $\mathrm{i}\hat{\gamma}_{-}^{\mu}\partial_{\mu}\Psi_{+}=\mathrm{i}\hat{\gamma}_{-}^{\mu}\partial_{\mu}(\Psi_{+})_{C}=0$
($\Rightarrow\mathrm{i}\hat{\gamma}^{\mu}\partial_{\mu}\Psi_{+}=\mathrm{i}\hat{\gamma}^{\mu}\partial_{\mu}(\Psi_{+})_{C}=0$). 

The Majorana condition imposed upon the two-component wave function
$\Psi$ gives us the following relations:
\begin{equation}
\Psi_{+}=(\Psi_{+})_{C}\quad\mathrm{and}\quad\Psi_{-}=(\Psi_{-})_{C},
\end{equation}
i.e., $\Psi_{+}=(\Psi_{C})_{+}$ and $\Psi_{-}=(\Psi_{C})_{-}$. Thus,
unlike what happens in (3+1) dimensions, $\Psi_{+}$ and $\Psi_{-}$
satisfy the Majorana condition. Clearly, the equation that describes
a Majorana particle in (1+1) dimensions is the pair of Eqs. (71) and
(72) (with the matrix relations (73)) and the pair of relations, or
restrictions, in (75) (the Majorana condition). Naturally, by imposing
the latter condition upon the equations in (74), we again obtain Eqs.
(71) and (72). 

On the other hand, making $\mathrm{m}c^{2}=V_{\mathrm{S}}=0$ in Eq.
(71) leads us to the relation $\mathrm{i}\hat{\gamma}_{+}^{\mu}\partial_{\mu}\Psi_{-}=0$
($\Rightarrow\mathrm{i}\hat{\gamma}^{\mu}\partial_{\mu}\Psi_{-}=0$),
and as can be seen in Eq. (74), we also have $\mathrm{i}\hat{\gamma}_{+}^{\mu}\partial_{\mu}(\Psi_{-})_{C}=0$
($\Rightarrow\mathrm{i}\hat{\gamma}^{\mu}\partial_{\mu}(\Psi_{-})_{C}=0$);
in this case also, we have $\Psi_{-}=(\Psi_{-})_{C}$ (this is due
to the Majorana condition). Similarly, making $\mathrm{m}c^{2}=V_{\mathrm{S}}=0$
in Eq. (72) leads us to the relation $\mathrm{i}\hat{\gamma}_{-}^{\mu}\partial_{\mu}\Psi_{+}=0$
($\Rightarrow\mathrm{i}\hat{\gamma}^{\mu}\partial_{\mu}\Psi_{+}=0$),
but from Eq. (74) we also have $\mathrm{i}\hat{\gamma}_{-}^{\mu}\partial_{\mu}(\Psi_{+})_{C}=0$
($\Rightarrow\mathrm{i}\hat{\gamma}^{\mu}\partial_{\mu}(\Psi_{+})_{C}=0$),
where $\Psi_{+}=(\Psi_{+})_{C}$ (because of the Majorana condition). 

Thus, to obtain the two-component wave function that describes the
one-dimensional Majorana particle, $\Psi=\Psi_{+}+\Psi_{-}$, we must
solve the system of equations formed by Eqs. (71) and (72), but $\Psi_{+}$
and $\Psi_{-}$ must verify the relations in Eq. (75), i.e., the Majorana
condition. Note that $\Psi=\Psi_{+}+\Psi_{-}=(\Psi_{+})_{C}+(\Psi_{-})_{C}$,
and therefore, $\Psi=\Psi_{C}$, as expected.

We can prove the following results. We make full use of Table 6. First,
in the Weyl representation, the covariant Eq. (71) for the two-component
wave functions $\Psi_{+}=\left[\,\varphi_{1}\;\,0\,\right]^{\mathrm{T}}$
and $\Psi_{-}=\left[\,0\;\,\varphi_{2}\,\right]^{\mathrm{T}}$ leads
us only to an equation for the one-component wave functions $\varphi_{1}$
and $\varphi_{2}$, namely, 
\begin{equation}
\mathrm{i}\hbar\frac{\partial}{\partial t}\varphi_{2}=+\mathrm{i}\hbar c\frac{\partial}{\partial x}\varphi_{2}+(V_{\mathrm{S}}+\mathrm{m}c^{2})\varphi_{1},
\end{equation}
and similarly, the covariant equation (72) leads us to 
\begin{equation}
\mathrm{i}\hbar\frac{\partial}{\partial t}\varphi_{1}=-\mathrm{i}\hbar c\frac{\partial}{\partial x}\varphi_{1}+(V_{\mathrm{S}}+\mathrm{m}c^{2})\varphi_{2}.
\end{equation}
The latter pair of equations comprises a complex system of coupled
equations; it is just Eq. (33), as expected. Likewise, the Majorana
condition in Eq. (75) leads us to the pair of relations in Eq. (34),
respectively, also as expected. Thus, we do not have a first-order
equation for a single component of the wave function in the Weyl representation.
Clearly, the four real degrees of freedom present in the solutions
of Eqs. (76) and (77) are reduced to only two due to the two relations
that emerge from the Majorana condition.

Incidentally, in (1+1) dimensions, one also has that $\varphi_{1}$
and $\varphi_{2}$ are transformed in two different ways under a Lorentz
boost. In effect, let us write the Lorentz boost along the $x$-axis
in the following way: $\left[\, ct'\; x'\,\right]^{\mathrm{T}}=\exp(-\omega\hat{\sigma}_{x})\left[\, ct\; x\,\right]^{\mathrm{T}}$
(i.e., $x^{\mu}\,'=\Lambda_{\;\,\nu}^{\mu}\, x^{\nu}$), where, as
usual, $\tanh(\omega)=v/c\equiv\beta$ and $\cosh(\omega)=(1-\beta^{2})^{-1/2}\equiv\gamma$,
with the speed of the primed (inertial) reference frame with respect
to the unprimed (inertial) reference frame being $v$. Then, under
this Lorentz boost, the wave function transforms as $\Psi'(x',t')=\hat{S}(\Lambda)\Psi(x,t)$,
where $\hat{S}(\Lambda)=\exp(-\omega\hat{\Gamma}^{5}/2)$ and which
obeys the relation $\Lambda_{\;\,\nu}^{\mu}\hat{\gamma}^{\nu}=\hat{S}^{-1}(\Lambda)\hat{\gamma}^{\mu}\hat{S}(\Lambda)$.
Then, just in the Weyl representation, the matrix $\hat{S}(\Lambda)$
is a diagonal matrix, and we obtain the following results:
\begin{equation}
\varphi_{1}'(x',t')=\left[\,\cosh\left(\frac{\omega}{2}\right)-\sinh\left(\frac{\omega}{2}\right)\right]\varphi_{1}(x,t)\,,\;\,\varphi_{2}'(x',t')=\left[\,\cosh\left(\frac{\omega}{2}\right)+\sinh\left(\frac{\omega}{2}\right)\right]\varphi_{2}(x,t).
\end{equation}
Thus, we have two different kinds of one-component wave functions
in (1+1) dimensions. Certainly, not only do $\varphi_{1}$ and $\varphi_{2}$
satisfy the relations in (78) but also $(\varphi_{1})_{C}$ and $(\varphi_{2})_{C}$.
This is because $\Psi$ and $\Psi_{C}$ are similarly transformed
under the Lorentz boost (i.e., $\Psi_{C}'(x',t')=\hat{S}(\Lambda)\Psi_{C}(x,t)$).
Interestingly, in the case where $\mathrm{m}c^{2}=V_{\mathrm{S}}=0$,
the wave functions with definite chirality, $\Psi_{+}$ and $\Psi_{-}$,
each satisfy the one-dimensional Dirac equation and their own Majorana
conditions. Also, in the Weyl representation, the nonzero component
of each of these two chiral wave functions satisfies a Weyl equation
(see Eqs. (76) and (77)). Thus, we could call the particles described
by $\Psi_{+}$ and $\Psi_{-}$ Weyl-Majorana particles \cite{RefY}. 

Second, in the Dirac representation, Eq. (71) leads us to Eq. (76)
and Eq. (72) leads us to Eq. (77) with the following replacements:
$\varphi_{1}\rightarrow\varphi+\chi$ and $\varphi_{2}\rightarrow\varphi-\chi$.
Likewise, the Majorana condition (Eq. (75)) leads us precisely to
the pair of relations in Eq. (34) with the latter replacements, namely,
$\varphi+\chi=-\mathrm{i}(\varphi+\chi)^{*}$ and $\varphi-\chi=+\mathrm{i}(\varphi-\chi)^{*}$
(the latter two relations imply the result given in Eq.(23)). Remember
that the two-component wave functions in the Dirac and Weyl representations
are related through the relation $\left[\,\varphi_{1}\;\varphi_{2}\,\right]^{\mathrm{T}}=\hat{S}\left[\,\varphi\;\chi\,\right]^{\mathrm{T}}$,
where the matrix $\hat{S}$ is given in Eq. (15). Certainly, adding
and subtracting the two equations obtained here and conveniently using
the Majorana condition again, we obtain an equation for the component
$\varphi$ of the wave function, namely, Eq. (24), and an equation
for the component $\chi$ of the wave function, namely, the same Eq.
(24) but with the following replacements: $\varphi\rightarrow\chi$
and $V_{\mathrm{S}}+\mathrm{m}c^{2}\rightarrow-(V_{\mathrm{S}}+\mathrm{m}c^{2})$.
As explained before, it is sufficient to solve only one of these two
equations because the remaining respective component can be obtained
from the Majorana condition. Thus, only two real quantities, or real
degrees of freedom, are sufficient to fully describe the Majorana
particle.

Third, in the Majorana representation, Eq. (71) leads us to Eq. (76)
and Eq. (72) leads us to Eq. (77) with the following replacements:
$\varphi_{1}\rightarrow(1-\mathrm{i})(\phi_{1}+\phi_{2})$ and $\varphi_{2}\rightarrow(1+\mathrm{i})(\phi_{1}-\phi_{2})$.
Remember that the two-component wave functions in the Majorana and
Weyl representations are related by $\left[\,\varphi_{1}\;\varphi_{2}\,\right]^{\mathrm{T}}=\hat{S}^{-1}\left[\,\phi_{1}\;\phi_{2}\,\right]^{\mathrm{T}}$,
where the matrix $\hat{S}$ is given in Eq. (14). In this representation,
$\hat{S}_{C}=\hat{1}_{2}$; therefore, the Majorana representation
(Eq. (75)) is simply $\Psi_{+}=\Psi_{+}^{*}$ and $\Psi_{-}=\Psi_{-}^{*}$,
i.e., the latter condition inmediately yields the pair of relations
$\phi_{1}+\phi_{2}=\phi_{1}^{*}+\phi_{2}^{*}$ and $\phi_{1}-\phi_{2}=\phi_{1}^{*}-\phi_{2}^{*}$,
respectively (which implies the result in Eq. (43), i.e., the entire
two-component wave function must be real). Finally, adding and subtracting
the two equations obtained here (but before multiplying by $\mathrm{i}(1-\mathrm{i})$
the equation that arises from Eq. (76) and multiplying by $\mathrm{i}(1+\mathrm{i})$
the one that emerges from Eq. (77)), we obtain a real system of coupled
equations, namely, the system in Eq. (44). Because the solutions of
this system are real-valued, the wave function has two real degrees
of freedom, as expected. 

\section{Conclusions}

\noindent Distinct differential equations can be used to describe
a Majorana particle in (3+1) and (1+1) dimensions. We can have a complex
single equation for a single component of the Dirac wave function,
as it is in the Dirac and Weyl representations in (3+1) dimensions
(in these cases, the single component itself is a two-component wave
function), and in the Dirac representation in (1+1) dimensions (in
this case, the single component itself is a one-component wave function).
Apropos of this, in the Weyl representation in (3+1) dimensions, one
can have two complex single equations, each being invariant under
its own type of Lorentz transformation (or Lorentz boost), i.e., these
two two-component covariant equations are non-equivalent equations,
and each of them can describe a specific type of Majorana particle
in (3+1) dimensions. Certainly, because of the Majorana condition,
the solutions of these two equations are not independent of each other,
that is, in the concrete description of the Majorana particle, two
plus two (complex) components are not absolutely necessary (i.e.,
the solution of only one of the two two-component Majorana equations
is what is needed to fully describe each type of Majorana particle).
Unexpectedly, in the Weyl representation in (1+1) dimensions, we have
a complex system of coupled equations, i.e., no first-order equation
for any of the components of the wave function can be written. On
the other hand, we can also have a real system of coupled equations,
as it is in the Majorana representation in (3+1) and (1+1) dimensions. 

All these equations and systems of equations emerge from the Dirac
equation and the Majorana condition when a representation is chosen.
Certainly, both the Dirac equation and the Majorana condition look
different written in their component forms when different representations
are used. In any case, whichever equation or system of equations is
used to describe the Majorana particle, the wave function that describes
it in (3+1) or (1+1) dimensions is determined by four or two real
quantities (real components, real and imaginary parts of complex components,
or just real or just imaginary parts of complex components), i.e.,
only four or two real quantities are sufficient.

Likewise, in (3+1) dimensions, the algebraic procedure introduced
by Case (and reexamined by us) allows us to write two covariant equations
of four components for the Majorana particle, i.e., in a form independent
of the choice of a particular representation for the matrices $\hat{\Gamma}^{\mu}$
and $\hat{\Lambda}^{\mu}$ (see Eqs. (53) and (55)). Each of these
equations provides one of the two covariant two-component Majorana
equations that arise when choosing the Weyl representation. In contrast,
in (1+1) dimensions, the algebraic procedure introduced by us leads
only to a covariant system of coupled first-order equations of two
components, and these components have their complex degrees of freedom
restricted by two conditions that arise from the Majorana condition.
This system of equations immediately gives us the complex system of
coupled first-order equations of one component that emerges when using
the Weyl representation, with the restriction given by the Majorana
condition. However, in the Dirac representation, the same system of
equations, together with the Majorana condition, can lead us to two
one-component equations (each for a single component of the two-component
wave function). 

It is hoped that our results can be useful to enrich the subject of
the distinct equations that can arise when describing the Majorana
particle in (1+1) and (3+1) dimensions. As we have seen, the results
obtained in these two space-time dimensions are not completely analogous.
It is to be expected that these results also present important differences
with results in (2+1) dimensions. However, in the latter case other
difficulties can also arise. Definitely, these issues should be treated
in another publication.

\begin{acknowledgments}
\noindent The author would like to thank Valedith Cusati, his wife,
for all her support. 
\end{acknowledgments}

\noindent \newpage{}

\noindent \begin{center}
\begin{tabular}{|c|c|c|c|c|c|}
\hline 
Representation & $\,\hat{\boldsymbol{\alpha}}\,$ & $\,\hat{\beta}\equiv\hat{\gamma}^{0}\,$ & $\,\hat{\beta}\hat{\boldsymbol{\alpha}}\equiv\hat{\boldsymbol{\gamma}}\,$ & $\hat{\gamma}^{5}\equiv\mathrm{i}\hat{\gamma}^{0}\hat{\gamma}^{1}\hat{\gamma}^{2}\hat{\gamma}^{3}$ & $\,\hat{S}_{C}=\hat{S}^{\dagger}\hat{S}^{*}\,$\tabularnewline
\hline 
\hline 
Dirac & $\hat{\sigma}_{x}\otimes\hat{\boldsymbol{\sigma}}$ & $\hat{\sigma}_{z}\otimes\hat{1}_{2}$ & $\mathrm{i}\hat{\sigma}_{y}\otimes\hat{\boldsymbol{\sigma}}$ & $\hat{\sigma}_{x}\otimes\hat{1}_{2}$ & $-\mathrm{i}\hat{\sigma}_{y}\otimes\hat{\sigma}_{y}$\tabularnewline
\hline 
Weyl & $\hat{\sigma}_{z}\otimes\hat{\boldsymbol{\sigma}}$ & $-\hat{\sigma}_{x}\otimes\hat{1}_{2}$ & $\mathrm{i}\hat{\sigma}_{y}\otimes\hat{\boldsymbol{\sigma}}$ & $\hat{\sigma}_{z}\otimes\hat{1}_{2}$ & $-\mathrm{i}\hat{\sigma}_{y}\otimes\hat{\sigma}_{y}$\tabularnewline
\hline 
Majorana & Table 1.1 & $\hat{\sigma}_{x}\otimes\hat{\sigma}_{y}$ & Table 1.2 & $\hat{\sigma}_{z}\otimes\hat{\sigma}_{y}$ & $\hat{1}_{2}\otimes\hat{1}_{2}$\tabularnewline
\hline 
\end{tabular}
\par\end{center}

\noindent \begin{flushright}
Table 1
\par\end{flushright}

\medskip{}

\noindent \begin{center}
\framebox{\begin{minipage}[t][1\totalheight][c]{0.19\columnwidth}%
\begin{center}
$\begin{array}{c}
\hat{\alpha}_{1}=-\hat{\sigma}_{x}\otimes\hat{\sigma}_{x}\\
\hat{\alpha}_{2}=\hat{\sigma}_{z}\otimes\hat{1}_{2}\\
\hat{\alpha}_{3}=-\hat{\sigma}_{x}\otimes\hat{\sigma}_{z}
\end{array}$
\par\end{center}%
\end{minipage}}
\par\end{center}

\noindent \begin{flushright}
Table 1.1
\par\end{flushright}

\medskip{}

\noindent \begin{center}
\framebox{\begin{minipage}[t][1\totalheight][c]{0.19\columnwidth}%
\begin{center}
$\begin{array}{c}
\hat{\gamma}^{1}=\mathrm{i}\hat{1}_{2}\otimes\hat{\sigma}_{z}\\
\hat{\gamma}^{2}=-\mathrm{i}\hat{\sigma}_{y}\otimes\hat{\sigma}_{y}\\
\hat{\gamma}^{3}=-\mathrm{i}\hat{1}_{2}\otimes\hat{\sigma}_{x}
\end{array}$
\par\end{center}%
\end{minipage}}
\par\end{center}

\noindent \begin{flushright}
Table 1.2
\par\end{flushright}

\medskip{}

\medskip{}

\medskip{}

\noindent \begin{center}
\begin{tabular}{|c|c|c|c|c|c|}
\hline 
Representation & $\,\hat{\alpha}\,$ & $\,\hat{\beta}\equiv\hat{\gamma}^{0}\,$ & $\,\hat{\beta}\hat{\alpha}\equiv\hat{\gamma}^{1}\,$ & $\hat{\Gamma}^{5}\equiv-\mathrm{i}\hat{\gamma}^{5}=\hat{\gamma}^{0}\hat{\gamma}^{1}$ & $\,\hat{S}_{C}=\hat{S}^{\dagger}\hat{S}^{*}\,$\tabularnewline
\hline 
\hline 
Dirac & $\hat{\sigma}_{x}$ & $\hat{\sigma}_{z}$ & $\mathrm{i}\hat{\sigma}_{y}$ & $\hat{\sigma}_{x}$ & $-\mathrm{i}\hat{\sigma}_{x}$\tabularnewline
\hline 
Weyl & $\hat{\sigma}_{z}$ & $\hat{\sigma}_{x}$ & $-\mathrm{i}\hat{\sigma}_{y}$ & $\hat{\sigma}_{z}$ & $-\mathrm{i}\hat{\sigma}_{z}$\tabularnewline
\hline 
Majorana & $\hat{\sigma}_{x}$ & $\hat{\sigma}_{y}$ & $-\mathrm{i}\hat{\sigma}_{z}$ & $\hat{\sigma}_{x}$ & $\hat{1}_{2}$\tabularnewline
\hline 
\end{tabular}
\par\end{center}

\noindent \begin{flushright}
Table 2
\par\end{flushright}

\noindent \newpage{}

\noindent \begin{center}
\begin{tabular}{|c|c|c|}
\hline 
Representation & $\,\Psi_{+}\,$ & $\,\Psi_{-}\,$\tabularnewline
\hline 
\hline 
Dirac & $\underset{}{\overset{}{\frac{1}{2}\left[\begin{array}{c}
\varphi+\chi\\
\varphi+\chi
\end{array}\right]}}$ & $\frac{1}{2}\left[\begin{array}{c}
\varphi-\chi\\
-\varphi+\chi
\end{array}\right]$\tabularnewline
\hline 
Weyl & $\overset{}{\underset{}{\left[\begin{array}{c}
\varphi_{1}\\
0
\end{array}\right]}}$ & $\left[\begin{array}{c}
0\\
\varphi_{2}
\end{array}\right]$\tabularnewline
\hline 
Majorana & $\quad$$\underset{}{\overset{}{\frac{1}{2}\left[\begin{array}{c}
(\hat{1}_{2}+\hat{\sigma}_{y})\phi_{1}\\
(\hat{1}_{2}-\hat{\sigma}_{y})\phi_{2}
\end{array}\right]}}$$\quad$ & $\quad$$\frac{1}{2}\left[\begin{array}{c}
(\hat{1}_{2}-\hat{\sigma}_{y})\phi_{1}\\
(\hat{1}_{2}+\hat{\sigma}_{y})\phi_{2}
\end{array}\right]$$\quad$\tabularnewline
\hline 
\end{tabular}
\par\end{center}

\noindent \begin{flushright}
Table 3
\par\end{flushright}

\noindent \newpage{}

\noindent \begin{center}
\begin{sideways}
\begin{tabular}{|c||c||c||c||c|}
\hline 
Representation & $\,\hat{\Gamma}^{0}\,$ & $\,\hat{\Gamma}^{1}\,$ & $\,\hat{\Gamma}^{2}\,$ & $\,\hat{\Gamma}^{3}\,$\tabularnewline
\hline 
\hline 
Dirac & $\underset{}{\overset{}{\frac{1}{2}\left[\begin{array}{cc}
-\hat{\sigma}_{y} & \;-\hat{\sigma}_{y}\\
-\hat{\sigma}_{y} & \;-\hat{\sigma}_{y}
\end{array}\right]}}$ & $\frac{1}{2}\left[\begin{array}{cc}
\mathrm{i}\hat{\sigma}_{z} & \;\mathrm{i}\hat{\sigma}_{z}\\
\mathrm{i}\hat{\sigma}_{z} & \;\mathrm{i}\hat{\sigma}_{z}
\end{array}\right]$ & $\frac{1}{2}\left[\begin{array}{cc}
-\hat{1}_{2} & \;-\hat{1}_{2}\\
-\hat{1}_{2} & \;-\hat{1}_{2}
\end{array}\right]$ & $\frac{1}{2}\left[\begin{array}{cc}
-\mathrm{i}\hat{\sigma}_{x} & \;-\mathrm{i}\hat{\sigma}_{x}\\
-\mathrm{i}\hat{\sigma}_{x} & \;-\mathrm{i}\hat{\sigma}_{x}
\end{array}\right]$\tabularnewline
\hline 
\hline 
Weyl & $\overset{}{\underset{}{\left[\begin{array}{cc}
-\hat{\sigma}_{y} & \;\hat{0}_{2}\\
\hat{0}_{2} & \;\hat{0}_{2}
\end{array}\right]}}$ & $\left[\begin{array}{cc}
\mathrm{i}\hat{\sigma}_{z} & \;\hat{0}_{2}\\
\hat{0}_{2} & \;\hat{0}_{2}
\end{array}\right]$ & $\left[\begin{array}{cc}
-\hat{1}_{2} & \;\hat{0}_{2}\\
\hat{0}_{2} & \;\hat{0}_{2}
\end{array}\right]$ & $\left[\begin{array}{cc}
-\mathrm{i}\hat{\sigma}_{x} & \;\hat{0}_{2}\\
\hat{0}_{2} & \;\hat{0}_{2}
\end{array}\right]$\tabularnewline
\hline 
\hline 
Majorana & $\;$$\underset{}{\overset{}{\frac{1}{2}\left[\begin{array}{cc}
\hat{0}_{2} & \;\hat{\sigma}_{y}-\hat{1}_{2}\\
\hat{\sigma}_{y}+\hat{1}_{2} & \;\hat{0}_{2}
\end{array}\right]}}$$\;$ & $\;$$\frac{1}{2}\left[\begin{array}{cc}
\mathrm{i}\hat{\sigma}_{z}+\hat{\sigma}_{x} & \;\hat{0}_{2}\\
\hat{0}_{2} & \;\mathrm{i}\hat{\sigma}_{z}-\hat{\sigma}_{x}
\end{array}\right]$$\;$ & $\;$$\frac{1}{2}\left[\begin{array}{cc}
\hat{0}_{2} & \;-\hat{\sigma}_{y}+\hat{1}_{2}\\
\hat{\sigma}_{y}+\hat{1}_{2} & \;\hat{0}_{2}
\end{array}\right]$$\;$ & $\;$$\frac{1}{2}\left[\begin{array}{cc}
-\mathrm{i}\hat{\sigma}_{x}+\hat{\sigma}_{z} & \;\hat{0}_{2}\\
\hat{0}_{2} & \;-\mathrm{i}\hat{\sigma}_{x}-\hat{\sigma}_{z}
\end{array}\right]$$\;$\tabularnewline
\hline 
\end{tabular}
\end{sideways}
\par\end{center}

\noindent \begin{flushright}
Table 4
\par\end{flushright}

\noindent \newpage{}

\noindent \begin{center}
\begin{sideways}
\begin{tabular}{|c||c||c||c||c|}
\hline 
Representation & $\,\hat{\Lambda}^{0}\,$ & $\,\hat{\Lambda}^{1}\,$ & $\,\hat{\Lambda}^{2}\,$ & $\,\hat{\Lambda}^{3}\,$\tabularnewline
\hline 
\hline 
Dirac & $\underset{}{\overset{}{\frac{1}{2}\left[\begin{array}{cc}
\hat{\sigma}_{y} & \;-\hat{\sigma}_{y}\\
-\hat{\sigma}_{y} & \;\hat{\sigma}_{y}
\end{array}\right]}}$ & $\frac{1}{2}\left[\begin{array}{cc}
\mathrm{i}\hat{\sigma}_{z} & \;-\mathrm{i}\hat{\sigma}_{z}\\
-\mathrm{i}\hat{\sigma}_{z} & \;\mathrm{i}\hat{\sigma}_{z}
\end{array}\right]$ & $\frac{1}{2}\left[\begin{array}{cc}
-\hat{1}_{2} & \;\hat{1}_{2}\\
\hat{1}_{2} & \;-\hat{1}_{2}
\end{array}\right]$ & $\frac{1}{2}\left[\begin{array}{cc}
-\mathrm{i}\hat{\sigma}_{x} & \;\mathrm{i}\hat{\sigma}_{x}\\
\mathrm{i}\hat{\sigma}_{x} & \;-\mathrm{i}\hat{\sigma}_{x}
\end{array}\right]$\tabularnewline
\hline 
\hline 
Weyl & $\overset{}{\underset{}{\left[\begin{array}{cc}
\hat{0}_{2} & \;\hat{0}_{2}\\
\hat{0}_{2} & \;\hat{\sigma}_{y}
\end{array}\right]}}$ & $\left[\begin{array}{cc}
\hat{0}_{2} & \;\hat{0}_{2}\\
\hat{0}_{2} & \;\mathrm{i}\hat{\sigma}_{z}
\end{array}\right]$ & $\left[\begin{array}{cc}
\hat{0}_{2} & \;\hat{0}_{2}\\
\hat{0}_{2} & \;-\hat{1}_{2}
\end{array}\right]$ & $\left[\begin{array}{cc}
\hat{0}_{2} & \;\hat{0}_{2}\\
\hat{0}_{2} & \;-\mathrm{i}\hat{\sigma}_{x}
\end{array}\right]$\tabularnewline
\hline 
\hline 
Majorana & $\;$$\underset{}{\overset{}{\frac{1}{2}\left[\begin{array}{cc}
\hat{0}_{2} & \;\hat{\sigma}_{y}+\hat{1}_{2}\\
\hat{\sigma}_{y}-\hat{1}_{2} & \;\hat{0}_{2}
\end{array}\right]}}$$\;$ & $\;$$\frac{1}{2}\left[\begin{array}{cc}
\mathrm{i}\hat{\sigma}_{z}-\hat{\sigma}_{x} & \;\hat{0}_{2}\\
\hat{0}_{2} & \;\mathrm{i}\hat{\sigma}_{z}+\hat{\sigma}_{x}
\end{array}\right]$$\;$ & $\;$$\frac{1}{2}\left[\begin{array}{cc}
\hat{0}_{2} & \;-\hat{\sigma}_{y}-\hat{1}_{2}\\
\hat{\sigma}_{y}-\hat{1}_{2} & \;\hat{0}_{2}
\end{array}\right]$$\;$ & $\;$$\frac{1}{2}\left[\begin{array}{cc}
-\mathrm{i}\hat{\sigma}_{x}-\hat{\sigma}_{z} & \;\hat{0}_{2}\\
\hat{0}_{2} & \;-\mathrm{i}\hat{\sigma}_{x}+\hat{\sigma}_{z}
\end{array}\right]$$\;$\tabularnewline
\hline 
\end{tabular}
\end{sideways}
\par\end{center}

\noindent \begin{flushright}
Table 5
\par\end{flushright}

\noindent \newpage{}

\noindent \begin{center}
\begin{tabular}{|c|c|c|c|c|}
\hline 
Representation & $\,\Psi_{+}\,$ & $\,\Psi_{-}\,$ & $\,\hat{\gamma}_{+}^{0}(=-\hat{\gamma}_{+}^{1})\,$ & $\,\hat{\gamma}_{-}^{0}(=\hat{\gamma}_{-}^{1})\,$\tabularnewline
\hline 
\hline 
Dirac & $\underset{}{\overset{}{\frac{1}{2}\left[\begin{array}{c}
\varphi+\chi\\
\varphi+\chi
\end{array}\right]}}$ & $\frac{1}{2}\left[\begin{array}{c}
\varphi-\chi\\
-\varphi+\chi
\end{array}\right]$ & $\frac{1}{2}\left[\begin{array}{cc}
1 & \;-1\\
1 & \;-1
\end{array}\right]$ & $\frac{1}{2}\left[\begin{array}{cc}
1 & \;1\\
-1 & \;-1
\end{array}\right]$\tabularnewline
\hline 
Weyl & $\overset{}{\underset{}{\left[\begin{array}{c}
\varphi_{1}\\
0
\end{array}\right]}}$ & $\left[\begin{array}{c}
0\\
\varphi_{2}
\end{array}\right]$ & $\left[\begin{array}{cc}
0 & \;1\\
0 & \;0
\end{array}\right]$ & $\left[\begin{array}{cc}
0 & \;0\\
1 & \;0
\end{array}\right]$\tabularnewline
\hline 
Majorana & $\quad$$\underset{}{\overset{}{\frac{1}{2}\left[\begin{array}{c}
\phi_{1}+\phi_{2}\\
\phi_{1}+\phi_{2}
\end{array}\right]}}$$\quad$ & $\quad$$\frac{1}{2}\left[\begin{array}{c}
\phi_{1}-\phi_{2}\\
-\phi_{1}+\phi_{2}
\end{array}\right]$$\quad$ & $\quad$$\frac{\mathrm{i}}{2}\left[\begin{array}{cc}
1 & \;-1\\
1 & \;-1
\end{array}\right]$$\quad$ & $\quad$$\frac{\mathrm{i}}{2}\left[\begin{array}{cc}
-1 & \;-1\\
1 & \;1
\end{array}\right]$$\quad$\tabularnewline
\hline 
\end{tabular}
\par\end{center}

\noindent \begin{flushright}
Table 6
\par\end{flushright}

\noindent \newpage{}


\begin{thebibliography}{10}
\bibitem{RefA}
E. Majorana, \textquotedblleft{}Teoria simmetrica dell\textquoteright{}elettrone
e del positrone\textquotedblright{}, Il Nuovo Cimento \textbf{14}
171-84 (1937).

\bibitem{RefB}
S. Esposito, \textquotedblleft{}Searching for an equation:
Dirac, Majorana and the others\textquotedblright{}, Ann. Phys. \textbf{327},
1617-44 (2012).

\bibitem{RefC}
P. B. Pal, \textquotedblleft{}Dirac, Majorana, and
Weyl fermions\textquotedblright{}, Am. J. Phys. \textbf{79}, 485-98
(2011).

\bibitem{RefD}
S. R. Elliott and M. Franz, ``Colloquium: Majorana
fermions in nuclear, particle, and solid-state physics'', Rev. Mod.
Phys. \textbf{87}, 137-63 (2015).

\bibitem[5]{RefE}
R. Aguado, \textquotedblleft{}Majorana quasiparticles
in condensed matter\textquotedblright{}, Rivista del Nuovo Cimento
\textbf{40}, 523-93 (2017).

\bibitem[6]{RefF}
S. De Vincenzo and C. S\'{a}nchez, \textquotedblleft{}General
boundary conditions for a Majorana single-particle in a box in (1+1)
dimensions\textquotedblright{}, Physics of Particles and Nuclei Letters
\textbf{15}, 257-68 (2018).

\bibitem[7]{RefG}
R. Keil, et al, \textquotedblleft{}Optical simulation
of charge conservation violation and Majorana dynamics\textquotedblright{},
Optica \textbf{2}, 454-9 (2015). 

\bibitem[8]{RefH}
K. M. Case, ``Reformulation of the Majorana theory
of the neutrino'', Phys. Rev. \textbf{107}, 307-16 (1957).

\bibitem[9]{RefI}
A. Aste, ``A direct road to Majorana fields'',
Symmetry \textbf{2}, 1776-809 (2010). 

\bibitem[10]{RefJ}
A. Zee, \textit{Quantum Field Theory in a Nutshell},
2nd ed. (Princeton University Press, Princeton, 2010).

\bibitem[11]{RefK}
J. J. Sakuray, \textit{Advanced Quantum Mechanics}
(Addison-Wesley, Reading, 1967). 

\bibitem[12]{RefL}
A. Messiah, \textit{Quantum Mechanics}, Vol. II
(North-Holland, Amsterdam, 1966). 

\bibitem[13]{RefM}
W-H. Steeb, \textit{Problems in Theoretical Physics},
Vol. II (BI-Wissenschaftsverlag, Mannhein, 1990); H. V. Henderson,
F. Pukelsheim and S. R. Searle, ``On the history of the Kronecker
product'', Linear and Multilinear Algebra \textbf{14}, 113-20 (1983).

\bibitem[14]{RefN}
M. H. Al-Hashimi, A. M. Shalaby and U. -J. Wiese,
``Majorana fermions in a box'', Phys. Rev. D \textbf{95}, 065007
(2017).

\bibitem[15]{RefO}
V. Alonso, S. De Vincenzo and L. Mondino, ``On
the boundary conditions for the Dirac equation'', Eur. J. Phys. \textbf{18},
315-20 (1997). 

\bibitem[16]{RefP}
W. Greiner, \textit{Relativistic Quantum Mechanics:
Wave Equations} (Springer, Berlin, 2000). 

\bibitem[17]{RefQ}
L. H. Ryder, \textit{Quantum Field Theory}, 2nd
ed. (Cambridge University Press, Cambridge, 1996). 

\bibitem[18]{RefR}
S. De Vincenzo, ``On real solutions of the Dirac
equation for a one-dimensional Majorana particle'', Results in Physics
\textbf{15}, 102598 (2019).

\bibitem[19]{RefS}
C. Noh, B. M. Rodr\'{\i}guez-Lara and D. G. Angelakis,
``Proposal for realization of the Majorana equation in a tabletop
experiment'', Phys. Rev. A \textbf{87}, 040102(R) (2013).

\bibitem[20]{RefT}
E. Marsch, ``The two-component Majorana equation
- Novel derivations and known symmetries'', J. Mod. Phys. \textbf{2},
1109-14 (2011).

\bibitem[21]{RefU}
R. N. Mohapatra and P. B. Pal, \textit{Massive
Neutrinos in Physics and Astrophysics}, 3rd ed. (World Scientific,
Singapore, 2004).

\bibitem[22]{RefV}
Y. F. P\'{e}rez and C. J. Quimbay, ``Sistema relativista
de dos niveles y oscilaciones de neutrinos de Majorana'', Revista
Colombiana de F\'{\i}sica \textbf{44}, 185-92 (2012). {[}in Spanish{]}

\bibitem[23]{RefW}
D. M. Gitman and A. L. Shelepin, ``Fields on the
Poincar\'{e} group: arbitrary spin description and relativistic wave equations'',
Int. J. Theor. Phys. \textbf{40}, 603-84 (2001).

\bibitem[24]{RefX}
D. B. Kaplan, ``Chiral symmetry and lattice fermions'',
Preprint, arXiv:0912.2560v2 {[}hep-lat{]} (2012).

\bibitem[25]{RefY}
S. De Vincenzo, \textquotedblleft{}On the boundary
conditions for the 1D Weyl-Majorana particle in a box\textquotedblright{},
Acta Phys. Pol. B \textbf{51}, 2055-64 (2020).
\end{thebibliography}
\end{document}